\PassOptionsToPackage{table}{xcolor}
\documentclass{bluecardreport}

\usepackage{amsmath,amsfonts,bm}

\def\eqref#1{equation~\ref{#1}}

\def\1{\bm{1}}

\DeclareMathAlphabet{\mathsfit}{\encodingdefault}{\sfdefault}{m}{sl}
\SetMathAlphabet{\mathsfit}{bold}{\encodingdefault}{\sfdefault}{bx}{n}

\usepackage{amssymb}
\usepackage{needspace}
\usepackage{subcaption}
\usepackage{multirow}
\usepackage{makecell}
\usepackage{colortbl}
\usepackage{tabularx}
\usepackage{float}
\usepackage{adjustbox}
\usepackage{pifont}

\newcommand{\benchmarkname}{AffectSim}
\newcommand{\presentmark}{\checkmark}
\newcommand{\absentmark}{\ding{55}}
\newcommand{\pendingresult}{\textcolor{gray}{---}}
\definecolor{benchmarkgroupbg}{RGB}{240,244,249}
\reportname{AffectSim: A Controllable Interactive 3D Simulation Benchmark for Embodied Affective Perception}
\reportbrand{AffectSim}
\reportlogo{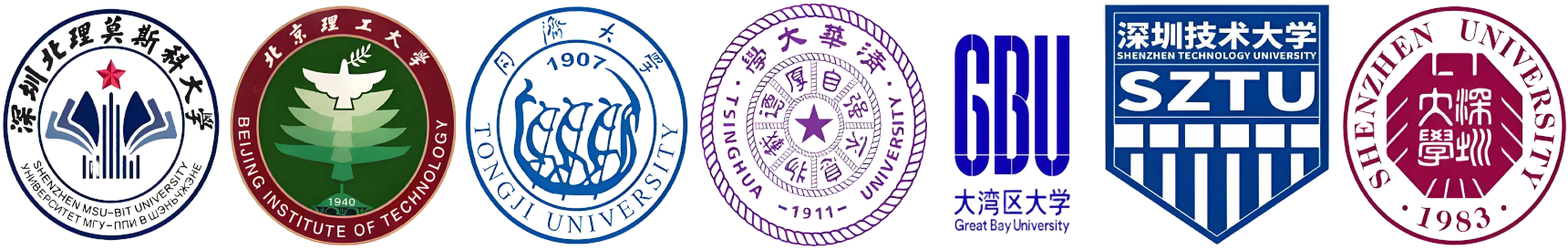}
\date{Technical Report --- September 28, 2026}
\title{AffectSim: A Controllable Interactive 3D Simulation Benchmark for Embodied Affective Perception}

\reportauthor{Ke Xing $^{1,2}$, Zhilong Wang$^{1,2}$, Zheng Lian$^3$, Sicheng Zhao$^4$, Haifeng Lu$^1$, Zhen Zhang$^1$,\\ Zitong Yu$^5$, Xiaojiang Peng$^6$, Changxin Huang$^7$, Runhao Zeng$^1$, Xiping Hu$^1$}

\reportaffiliation{$^1$Shenzhen MSU-BIT University,  
$^2$Beijing Institute of Technology, 
$^3$Tongji University, 
$^4$Tsinghua University, \\
$^5$Great Bay University, 
$^6$Shenzhen Technology University,
$^7$Shenzhen University}

\hypersetup{
  pdfauthor={Ke Xing, Zhilong Wang, Zheng Lian, Sicheng Zhao, Haifeng Lu, Zhen Zhang, Zitong Yu, Xiaojiang Peng, Changxin Huang, Runhao Zeng, Xiping Hu},
  pdfsubject={Embodied affective perception benchmark},
  pdfkeywords={AffectSim, embodied affective perception, active perception}
}

\reportabstract{Existing affective benchmarks largely rely on fixed recordings, where
observation conditions are predetermined before inference. Consequently, they
mainly evaluate passive affect recognition while overlooking a key question
for embodied agents: how to actively acquire informative affective evidence.
We introduce \textbf{\benchmarkname{}}, a controllable interactive 3D
simulation benchmark for embodied affective perception. \benchmarkname{}
represents affective behaviors as replayable 3D episodes and separates the
underlying behavior from how it is observed, enabling systematic control over
distance, orientation, occlusion, scene geometry, and agent viewpoint.
It contains 27,647 episodes across five emotion categories and 57 scenes.
Across 24 frozen perception-model configurations, we show that observation
quality substantially affects recognition. To examine whether active sensing
can recover this gap, we establish a training-free heuristic baseline,
which improves recognition in 21 of 24 settings. Building on this, we  propose
Evidence-Aware Observation Gate (EAOG) to adaptively assess whether
additional observation is still beneficial.
These results show that affective perception depends not only on the observed
behavior, but also on how evidence is acquired. By making observation
acquisition experimentally controllable, \benchmarkname{} provides a foundation
for studying embodied affective perception in interactive 3D environments. All benchmark assets, code, and simulation environments will be released.
}
\reportteaser{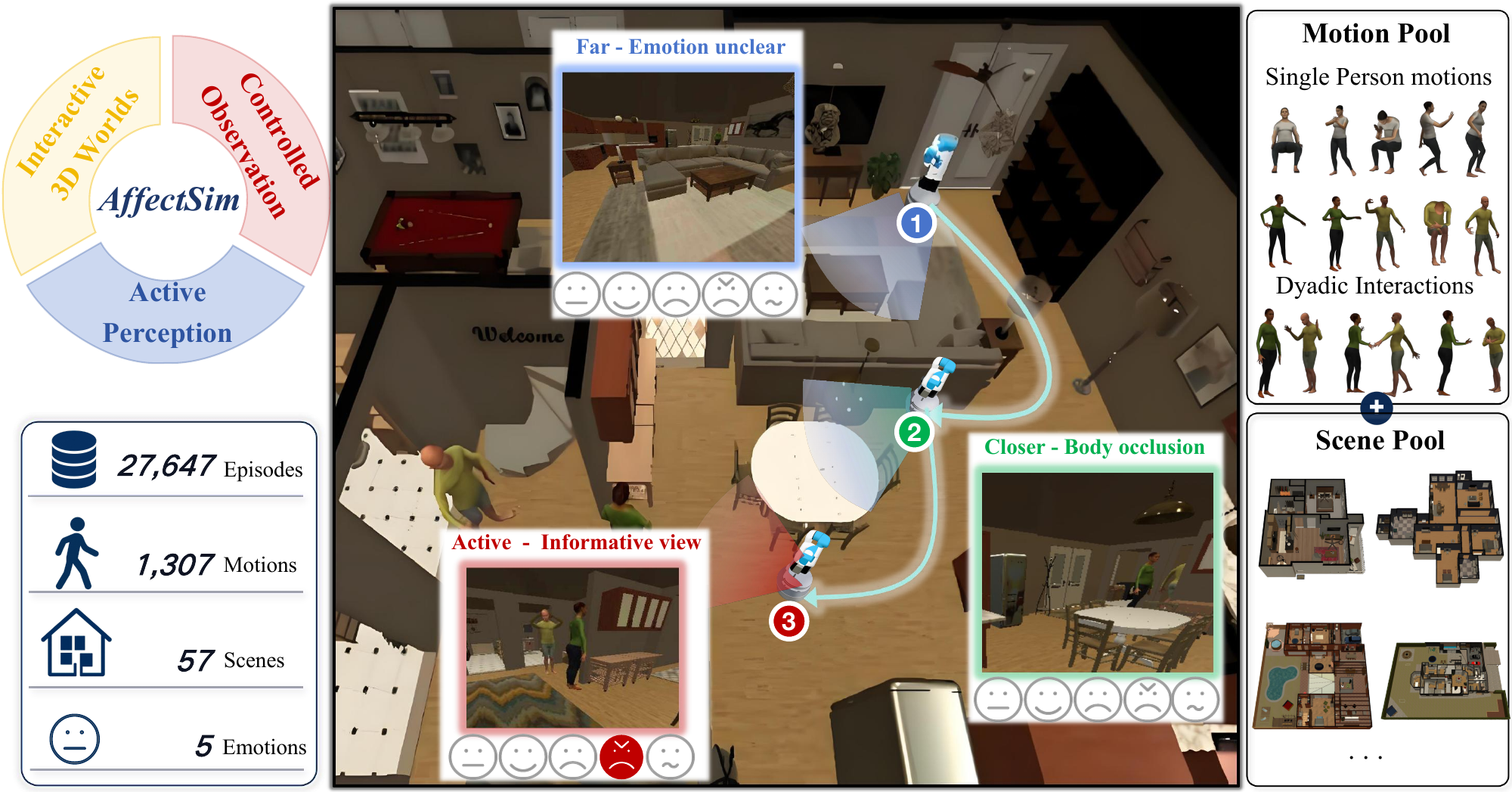}
\reportteasercaption[fig:shouye]{%
Overview of \benchmarkname{}. Reusable affective motions are instantiated in
  interactive 3D scenes under controllable observation conditions. In the
  illustrated episode, a distant view provides limited evidence, approaching
  remains insufficient under occlusion, and active repositioning reveals a
  more informative view.
}

\begin{document}
\maketitle
\clearpage
\reportpagestyle
\tableofcontents
\clearpage

\section{Introduction}
\label{sec:intro}

Robots operating around people must understand not only objects and
instructions, but also human affect~\citep{narayanan2020proxemo}. Yet most
emotion-recognition benchmarks evaluate observations that have already been
collected and framed before inference
~\citep{poria2019meld,wang2022ferv39k}. This implicitly
assumes an \emph{ideal photographer} who places the camera at an informative
distance and orientation and makes the relevant behavior visible. Such an
assumption does not hold for embodied agents: as illustrated in Figure
\ref{fig:shouye}, the same affective behavior may appear ambiguous from afar,
remain partially occluded after approaching, and only become informative after
active repositioning. For embodied affective perception, a good viewpoint is
not given; it must be acquired.

This limitation reflects the prevailing collect-annotate-recognize paradigm in
affective computing. Once affective behavior is recorded, viewpoint, distance,
orientation, occlusion, and framing become entangled with the sample itself.
Although substantial progress has been made in recognizing richer affective
signals~\citep{cheng2024emotion}, the same behavior cannot easily be
re-observed under systematically changed conditions. It is therefore difficult
to distinguish failures of affective understanding from failures of
observation.

Interactive 3D simulation makes this missing dimension experimentally
controllable~\citep{savva2019habitat}. By instantiating affective behaviors as
replayable events in executable worlds, the same behavior can be re-observed
under different conditions, while an embodied agent can actively change the
evidence available to it through action
~\citep{bajcsy1988active,yang2019embodiedamodal}.
Simulation here is not merely a means of scaling data, but an experimental
instrument for controlled and repeatable re-observation that is difficult to
realize in the physical world.

To this end, we introduce \benchmarkname{}, a controllable interactive 3D
benchmark for embodied affective perception. Its central design principle is
to separate \emph{what affective behavior is instantiated} from \emph{how it is
observed}. \benchmarkname{} contains 27{,}647 replayable episodes across 57
scenes, covering single-person and dyadic behaviors while independently
varying viewpoint, distance, orientation, occlusion, and scene configuration.
Matched passive and active observation protocols quantify the resulting
\emph{observation-quality gap} and how much of it can be recovered through
embodied sensing.

Our contributions are threefold:

\noindent\textbf{(1) A controllable benchmark for embodied affective perception.}
\benchmarkname{} enables matched re-observation of the same affective behavior
under systematically varied observation conditions and agent actions.

\noindent\textbf{(2) A systematic study of observation-driven affective perception.}
Experiments across 24 frozen recognizer configurations reveal a substantial
observation-quality gap and show that active sensing can partially recover it.

\noindent\textbf{(3) Reference solutions for active affective observation.}
We propose a training-free heuristic baseline and a learned closed-loop system
with Evidence-Aware Observation Gate (EAOG), providing initial
reference results for active evidence acquisition and its quality--cost
trade-off.

\section{Related Work}
\label{sec:related}

\begin{table}[t]
\centering
\setlength{\belowcaptionskip}{\baselineskip}
\caption{
Comparison of representative affective and embodied benchmarks.
\presentmark{} and \absentmark{} indicate whether each capability
is supported in an official evaluation setting.
}
\vspace{-0.3cm}
\label{tab:benchmark-comparison}

\begingroup
\renewcommand{\baselinestretch}{1}
\small
\setlength{\tabcolsep}{3pt}
\setlength{\extrarowheight}{0pt}
\renewcommand{\arraystretch}{1.05}
\setlength{\aboverulesep}{0.4ex}
\setlength{\belowrulesep}{0.4ex}

\newcommand{\benchhead}[1]{{\scriptsize\bfseries #1}}

\begin{tabularx}{\textwidth}{@{}
    >{\raggedright\arraybackslash}X
    >{\centering\arraybackslash}p{0.090\textwidth}
    >{\centering\arraybackslash}p{0.090\textwidth}
    >{\centering\arraybackslash}p{0.135\textwidth}
    >{\centering\arraybackslash}p{0.115\textwidth}
    >{\centering\arraybackslash}p{0.090\textwidth}
@{}}
\toprule

\multirow{2}{*}{\benchhead{Benchmark}}

& \benchhead{Egocentric}
& \benchhead{3D}
& \benchhead{Interactive}
& \benchhead{Controlled}
& \benchhead{Agent}
\\[-2pt]

& \benchhead{view}
& \benchhead{humans}
& \benchhead{environment}
& \benchhead{replay}
& \benchhead{control}
\\

\midrule

\rowcolor{benchmarkgroupbg}
\multicolumn{6}{@{}l@{}}{
    \strut\emph{Passive affective benchmarks}
}
\\

Emilya~\citep{fourati2014emilya}

& \absentmark & \absentmark & \absentmark
& \absentmark & \absentmark
\\

KDAE~\citep{zhang2020kinematic}

& \absentmark & \absentmark & \absentmark
& \absentmark & \absentmark
\\

DFEW~\citep{jiang2020dfew}

& \absentmark & \absentmark & \absentmark
& \absentmark & \absentmark
\\

iMiGUE~\citep{liu2021imigue}

& \absentmark & \absentmark & \absentmark
& \absentmark & \absentmark
\\

MER2024~\citep{lian2024mer2024}

& \absentmark & \absentmark & \absentmark
& \absentmark & \absentmark
\\

MERBench~\citep{lian2026merbench}

& \absentmark & \absentmark & \absentmark
& \absentmark & \absentmark
\\

\midrule

\rowcolor{benchmarkgroupbg}
\multicolumn{6}{@{}l@{}}{
    \strut\emph{Egocentric and social-affective benchmarks}
}
\\

$E^3$~\citep{lin2024e3}

& \presentmark & \absentmark & \absentmark
& \absentmark & \absentmark
\\

EmpathyAgent~\citep{chen2025empathyagent}

& \presentmark & \presentmark & \absentmark
& \absentmark & \absentmark
\\

OSMO~\citep{abdelfattah2026osmo}

& \presentmark & \absentmark & \absentmark
& \absentmark & \absentmark 
\\

RobotEQ~\citep{fang2026roboteq}

& \presentmark & \absentmark & \absentmark
& \absentmark & \absentmark
\\

\midrule

\rowcolor{black!6}
\textbf{\benchmarkname{} (ours)}

& \presentmark & \presentmark & \presentmark
& \presentmark & \presentmark
\\

\bottomrule
\end{tabularx}
\endgroup
\end{table}

\paragraph{Affective perception from fixed observations.}
Affective computing has progressively expanded from facial and conversational
cues to contextual, bodily, multimodal, and motion-level evidence
\citep{poria2019meld,kosti2017emotic,liu2021imigue,
fourati2014emilya,zhang2020kinematic}.
Recent foundation-model benchmarks further emphasize affective reasoning and
explanation
\citep{cheng2024emotion,lian2025affectgpt,zhang2026mmeemotion}.
Despite increasingly rich evidence and stronger models, these settings share a
common assumption: observations are collected and fixed before inference,
leaving how affective evidence is acquired largely outside the evaluation loop.

\paragraph{Embodied affective perception and active sensing.}
Robot-centric datasets, $E^3$, OSMO, RobotEQ, and EmpathyAgent move affective
understanding toward egocentric or embodied settings
\citep{quiroz2022robotcentric,lin2024e3,abdelfattah2026osmo,
fang2026roboteq,chen2025empathyagent}.
However, an egocentric viewpoint does not by itself make observation
controllable: the physical trajectory that determines the available evidence
is still largely predefined. Active perception instead couples sensing with
action so that an agent can change subsequent observations
\citep{bajcsy1988active,yang2019embodiedamodal,fan2024activeopenvocab}.
\benchmarkname{} brings this capability to affective perception by enabling
the same affective behavior to be replayed under controlled observation
conditions while allowing an embodied agent to actively acquire new evidence.
We summarize the key differences among representative benchmarks in Table \ref{tab:benchmark-comparison}.
A detailed comparison with existing benchmarks is provided in Section
\ref{app:benchmark-comparison}.

\section{\benchmarkname{} Benchmark}
\label{sec:benchmark}

\begin{figure}[t]
  \centering
  \includegraphics[width=\textwidth]{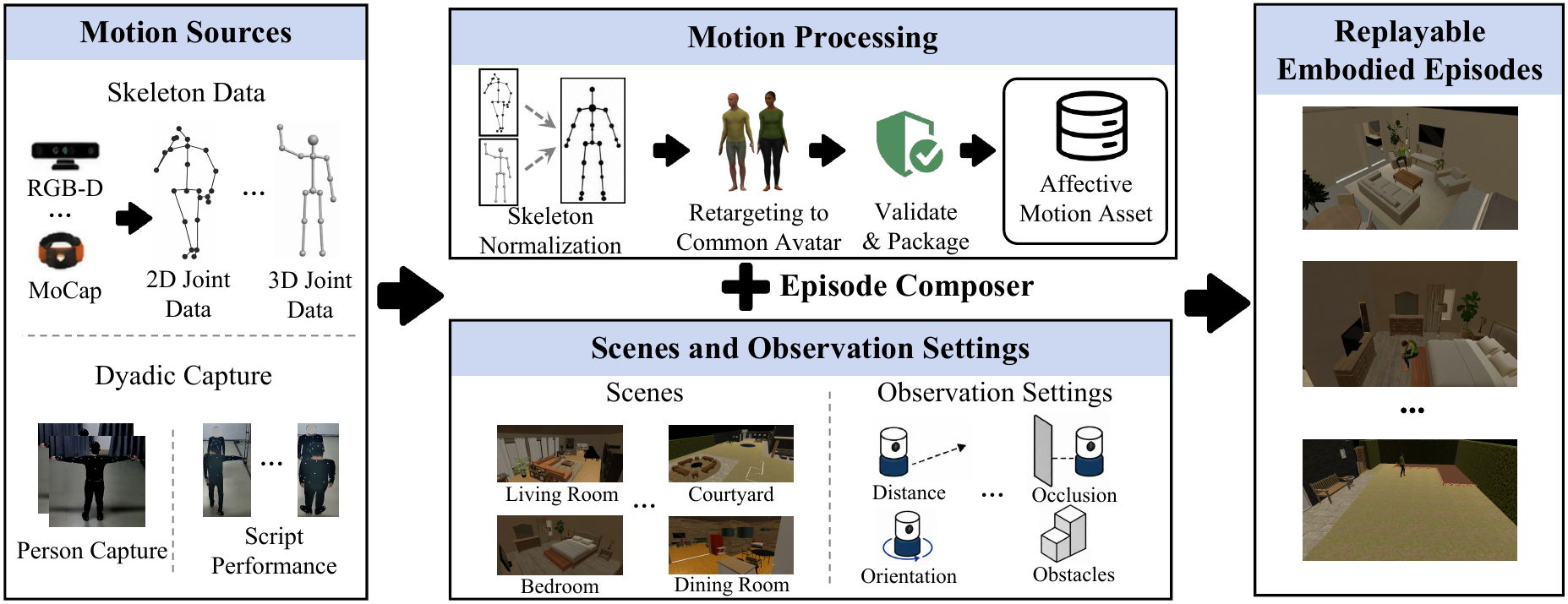}
  \vspace{-0.5cm}
  \caption{
  Construction pipeline of \benchmarkname{}. Emotion-labeled performances are
  converted into reusable motion assets and composed with scenes and controlled
  observation conditions to form replayable embodied episodes.
  }
  \label{fig:pipeline}
\end{figure}

\benchmarkname{} is built around a simple principle:
\emph{affective behavior and its observation conditions should be independently
controllable}. As illustrated in Figure \ref{fig:pipeline}, emotion-labeled
performances are converted into reusable motion assets and composed with
3D scenes, human placements, and observation challenges to form executable
episodes.

The benchmark is implemented in Habitat \citep{savva2019habitat} with HSSD
indoor scenes \citep{khanna2024hssd}. We represent an episode as
\begin{equation}
  e=\mathcal{C}(m,S,p_h,x_0),
  \label{eq:episode-composition}
\end{equation}
where $m$ is an affective motion, $S$ a scene, $p_h$ the human placement and
target identity, and $x_0$ the agent initialization. The emotion label is
inherited from $m$, while distance, orientation, occlusion, scene geometry,
and viewpoint are determined by the embodied configuration. Thus, the same
underlying behavior can be re-observed under different conditions rather than
being tied to a pre-rendered video.

\subsection{Affective Motion Assets}
\label{sec:curation}

The motion library combines KDAE~\citep{zhang2020kinematic},
Emilya~\citep{fourati2014emilya}, and a newly captured dyadic pilot dataset,
covering scenario-based, everyday-action, and interpersonal affective
behaviors. We harmonize their annotations into five shared classes:
\emph{angry}, \emph{fearful}, \emph{happy}, \emph{neutral}, and \emph{sad},
excluding categories without clear cross-dataset correspondence.
Motions that can be faithfully instantiated in simulation are retained and
retargeted to a common avatar while preserving their source labels and
provenance. A blind human validation on 500 randomly sampled retargeted
motions achieves 77.81\% Macro-F1 against the source emotion labels, supporting
that affective semantics are largely preserved after retargeting.
Detailed motion selection, retargeting, and validation procedures are provided
in Appendix~\ref{app:benchmark-details}.

\subsection{Embodied Episode Generation}
\label{sec:episodes}

The episode composer instantiates \cref{eq:episode-composition} using two
complementary construction modes according to whether the motion semantics
depend on the surrounding environment.

\noindent\textbf{Action-grounded composition.}
Motions such as sitting, door knocking, and walking are aligned with compatible
scene geometry and navigable trajectories to ensure physically and semantically
valid execution.

\noindent\textbf{Observation-controlled composition.}
For motions without object-specific dependencies, the affective behavior is
held fixed while the initial observation is varied. We construct four challenge
families---\emph{far-visible}, \emph{rear-view}, \emph{partial-occlusion}, and
\emph{obstacle-detour}---to stress target scale, viewing direction, body
visibility, and navigational accessibility. The same abstraction also supports
dyadic behavior.

\noindent\textbf{Episode execution.}
The robot is initialized on navigable floor space and the affective motion is
replayed cyclically during the episode. This factors out event-timing
uncertainty and isolates the spatial active-perception problem: whether an
agent can acquire a viewpoint that exposes discriminative bodily cues.
\subsection{Validation, Scale, and Splits}
\label{sec:validation}
\label{sec:statistics}
\label{sec:data-splits}

Generated episodes undergo automated checks for configuration integrity,
physical validity, reproducibility, and realization of the intended observation
challenge. A pool of 15 reviewers further screens candidate episodes for
physical plausibility, motion--scene compatibility, and rendering quality;
only episodes judged reasonable are retained.

The final benchmark contains 27{,}647 episodes across 57 scenes, with
19{,}662/4{,}044/3{,}941 episodes in the train/validation/test splits.
All variants derived from the same underlying performance are assigned to the
same split to prevent motion leakage. Scenes are shared across splits, so the
benchmark evaluates generalization to unseen performers and source motions
within a common scene inventory rather than to unseen environments.
Detailed split statistics, validation criteria, and initialization procedures are provided in  Section \ref{app:benchmark-details}.

\section{Evaluation Protocol}
\label{sec:protocol}

\subsection{Matched Observation Settings}
\label{sec:observation-settings}

All paired settings preserve the scene, target motion, motion phase, emotion
label, rendering parameters, and robot initialization. Within each experiment,
they also share the same temporal budget and recognition interface; only the
observation trajectory changes.

\noindent\textbf{P-Init (Initial View).}
The camera remains fixed at the agent's initial pose $x_0$, measuring the
evidence available before any embodied action.
\textbf{P-Ref (Reference View).}
A privileged reference observer maintains a reproducible informative view
according to body visibility, alignment, distance, framing, and scene
clearance. P-Ref is a diagnostic reference rather than an oracle or upper
bound.
\textbf{P-Mon (Monitor View).}
A fixed room-level camera provides an external monitoring view and serves as
an additional passive diagnostic.
\textbf{A-Obs (Active View).}
\label{sec:active-observation}
A-Obs starts from the same robot and human state as P-Init, but allows the
observation policy to change the robot viewpoint.

\subsection{Evaluation Metrics}
\label{sec:metrics}

\noindent\textbf{Recognition quality.}
We use Macro-F1 as the primary five-class recognition metric and additionally
report accuracy and probability-based recognition metrics where appropriate.

\textbf{Observation benefit.}
For a higher-is-better metric $M$ and active policy $\pi$, the paired gain is
$\Delta_{\mathrm{gain}}(M;\pi)
=M_{\pi}-M_{\mathrm{P\text{-}Init}}$.
To quantify how much of the observation-quality gap is recovered, we define
\begin{equation}
  \mathrm{OGR}(M;\pi)=
  \frac{M_{\pi}-M_{\mathrm{P\text{-}Init}}}
       {M_{\mathrm{P\text{-}Ref}}-M_{\mathrm{P\text{-}Init}}},
  \label{eq:ogr}
\end{equation}
when $M_{\mathrm{P\text{-}Ref}}>M_{\mathrm{P\text{-}Init}}$.
Because P-Ref is a reproducible reference rather than an oracle, OGR is not clipped to $[0,1]$.

\textbf{Recovery and acquisition cost.}
Recovery Rate measures how often active observation corrects an initially
incorrect prediction. We additionally report path length and Emotion-SPL
(E-SPL) as navigation-aware acquisition diagnostics following
\citet{anderson2018evaluationembodiednavigationagents}. Formal definitions
and supplementary execution diagnostics are provided in
Section \ref{app:protocol-details}.

\section{Active Affective Perception}
\label{sec:method}

\begin{figure}[t]
  \centering
  \includegraphics[width=\textwidth]{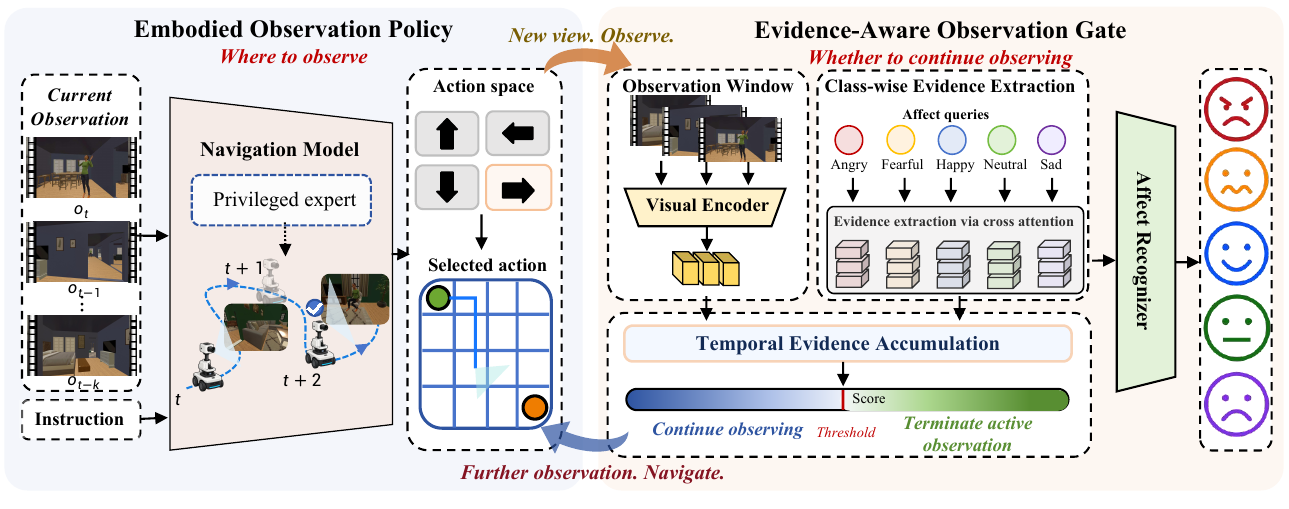}

  \caption{
  Overview of our closed-loop active affective perception framework.
  As the agent moves and observes, the observation policy controls viewpoint
  acquisition while EAOG continuously accumulates affective evidence and
  assesses whether further observation is beneficial.
  }
  \label{fig:method}
\end{figure}

We formulate embodied affective perception as a closed-loop active sensing problem, as illustrated in Figure \ref{fig:method}. At each step, the agent observes the target, adjusts its viewpoint,
and updates its accumulated affective evidence. The observation policy determines \emph{where to observe}, while the
Evidence-Aware Observation Gate (EAOG) evaluates the accumulated affective
evidence and determines whether continued observation remains worthwhile. Thus, viewpoint control and affective
evidence assessment proceed concurrently throughout interaction.

\subsection{Closed-Loop Active Observation}
\label{sec:active-observation-loop}

\paragraph{Embodied observation policy.}
We adapt Uni-NaVid
\citep{zhang2025uninavidvideobasedvisionlanguageactionmodel} with a movement
head over five actions: \textsc{Forward}, \textsc{Backward}, \textsc{Left},
\textsc{Right}, and \textsc{Wait}. At each step, the policy receives recent
egocentric RGB observations, an instruction to maintain an informative view
of the target person, and its executed-action history, and predicts the next
movement action.

\paragraph{Evidence-aware observation termination.}
In parallel with viewpoint control, EAOG continuously evaluates the affective
evidence acquired so far. At each decision time $t\in\mathcal T$, it extracts
spatiotemporal features from the current RGB window $\mathcal W_t$ and updates
emotion-conditioned evidence memories:
\begin{equation}
\mathbf z_{t,c}
=
\operatorname{Attend}
(\mathbf q_c,\operatorname{Enc}(\mathcal W_t)),
\qquad
\mathbf e_{t,c}
=
\operatorname{GRU}
(\mathbf z_{t,c},\mathbf e_{t-1,c}).
\label{eq:method-evidence}
\end{equation}
The accumulated evidence states produce a termination probability:
\begin{equation}
h_t=
\sigma\!\left(
g_{\mathrm{term}}
([\mathbf e_{t,1};\ldots;\mathbf e_{t,C}])
\right).
\label{eq:method-hazard}
\end{equation}
Here, $h_t$ measures whether further sensing is still worthwhile rather than
emotion confidence itself. If the accumulated evidence remains insufficient,
the agent continues moving and observing; otherwise, active observation
terminates. The resulting system therefore follows a continuous
\emph{move--observe--update--assess} loop.

\subsection{Learning the Active Observation System}
\label{sec:active-observation-training}

\paragraph{Learning where to observe.}
We train the observation policy through behavior cloning from a privileged
teacher. The teacher uses simulator-side information, including scene
geometry, navigation paths, target motion, and body visibility, to generate
actions that improve affective evidence acquisition. The student learns these
actions using only observations available at inference.

\paragraph{Learning when to terminate observation.}
EAOG is trained from observation trajectories while the learned
observation policy and affect recognizer remain frozen. A frozen affect
recognizer evaluates each candidate window and provides a recognition risk
$R_t$ based on the Half-Brier score. We combine recognition risk with
observation delay:
\begin{equation}
C_t=R_t+\lambda(t-t_{\min}),
\qquad
d_t=
\frac{\exp(-C_t/\tau)}
{\sum_j\exp(-C_j/\tau)}.
\label{eq:time_loss}
\end{equation}
The soft target $d_t$ supervises when additional observation is no longer
worthwhile. Emotion classification and quality prediction provide auxiliary
supervision:
\begin{equation}
\mathcal L_{\mathrm{EAOG}}
=
\mathcal L_{\mathrm{term}}
+
\mathcal L_{\mathrm{quality}}
+
\mathcal L_{\mathrm{emotion}}.
\label{eq:method-eqt-objective}
\end{equation}
Detailed scoring, supervision, and
optimization are provided in Section \ref{app:eaot-supervision}.

\section{Experiments}
\label{sec:experiments}

\subsection{Experimental Setup}
\label{sec:exp-setup}

We evaluate all 3{,}941 test episodes under two complementary settings.
In the \emph{zero-shot setting}, 24 frozen affect recognizers
(19 open-source and 5 closed-source) compare P-Init, P-Mon, P-Ref, and a
training-free heuristic active baseline. The baseline uses a frozen planner to locate
the target, followed by a depth- and location-based controller to maintain an
informative view, without emotion supervision or recognizer feedback. If no
valid new observation is acquired, it falls back to the paired P-Init view.

In the \emph{learned setting}, we evaluate EAOG with a behavior-cloned
observation policy and an AffectSim-adapted frozen VideoLLaMA2.1-7B-AV
recognizer. Controlled comparisons fix the navigator, recognizer, and candidate
observations, varying only the observation-gating strategy. Implementation and inference details are provided in
Sections~\ref{app:training-details} and~\ref{app:eaot-implementation}.

\begin{table}[t]
\centering
\setlength{\belowcaptionskip}{\baselineskip}
\caption{
Mean test Macro-F1 (\%) across 24 frozen affect recognizers under matched
observation conditions. P-Init, P-Mon, and P-Ref denote passive observations;
Heuristic denotes the training-free active-observation baseline, with P-Init
fallback when no valid new observation is acquired. Full per-recognizer results
are reported in Appendix Table~\ref{app:tab:main-results-full}.
}

\label{tab:main-results}
\small
\setlength{\tabcolsep}{3pt}
\renewcommand{\arraystretch}{1.15}
\begin{tabular}{l c l c c c | c c}
\toprule
\textbf{Recognizer group} & \textbf{\# Models} & \textbf{Examples}
& \textbf{P-Mon} & \textbf{P-Ref} & \textbf{P-Init}
& \textbf{Heur.}
& \textbf{Heur.$-$P-Init}\\
\midrule
Open-source   & 19 & Qwen, AffectGPT      & 10.31 & 15.56 & 9.89  & 11.70 & +1.81 \\
Closed-source & 5  & GPT, Gemini, Claude & 24.44 & 30.78 & 22.61 & 24.26 & +1.65 \\
\midrule
Overall       & 24 & --                  & 13.26 & 18.73 & 12.54 & 14.32 & +1.78 \\
\bottomrule
\end{tabular}
\end{table}

\subsection{Q1: Does Observation Quality Matter?}
\label{sec:exp-observation-quality}

We isolate the effect of observation quality by holding the affective motion,
scene, motion phase, temporal budget, and recognizer interface fixed for each
matched episode--recognizer pair, while varying only the observation condition.
We compare the initial egocentric view (P-Init), fixed room-monitor view
(P-Mon), and privileged reference view (P-Ref).

As shown in Table \ref{tab:main-results}, P-Ref substantially improves recognition
over P-Init across both open- and closed-source recognizers. Mean Macro-F1
increases from 9.89\% to 15.56\% for the 19 open-source configurations and
from 22.61\% to 30.78\% for the five closed-source configurations; P-Ref
achieves the best passive-view performance for 23 of 24 configurations.
This establishes a substantial \emph{observation-quality gap}: recognition
performance depends not only on the recognizer, but also on the evidence made
available to it.

The gain is not explained simply by replacing the egocentric view with an
external camera. P-Mon reaches only 10.31\% and 24.44\% Macro-F1 for the
open- and closed-source groups, respectively, remaining much closer to P-Init
than to P-Ref. Thus, informative observation depends on factors such as
visibility, distance, orientation, and framing rather than camera placement
alone. P-Ref is a reproducible diagnostic reference rather than an oracle or
upper bound. Complete per-recognizer results for all passive observation
conditions are reported in Table~\ref{app:tab:main-results-full}.

\begin{figure}[t]
  \centering
  \includegraphics[width=0.97\linewidth]{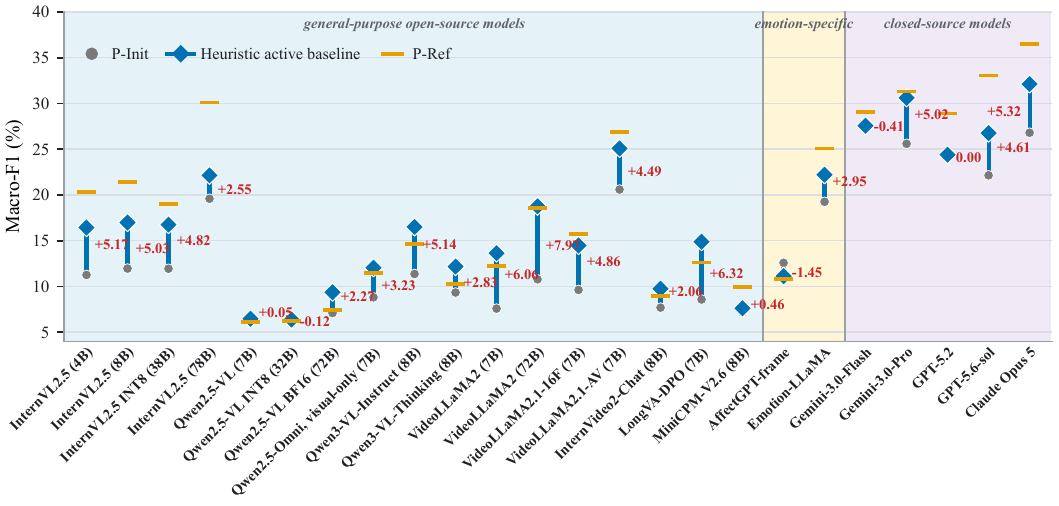}

  \caption{
  Paired Macro-F1 (\%) of 24 frozen affect recognizers on the 1{,}881
  test episodes where the heuristic baseline successfully acquires a valid
  new observation. Red labels show the
  paired gain over P-Init. This subset isolates the quality of newly acquired
  observations from acquisition failures.
  }
  \label{fig:new-video-gains}
\end{figure}

\begin{table}[t]
\centering
\caption{
Controlled comparison of observation strategies on test episodes
using the same AffectSim-adapted and frozen VideoLLaMA2.1-7B-AV recognizer.
P-Init is the reference for OGR.
}

\label{tab:system_comparison}
\small
\begin{tabular}{@{}llccc@{}}
\toprule
\textbf{Method} & \textbf{Observation strategy} & \textbf{F1$\uparrow$} & \textbf{OGR$\uparrow$} & \textbf{E-SPL@3$\uparrow$} \\
\midrule
P-Init & Fixed initial view & 32.74 & -- &33.80 \\
Heuristic & Hand-designed active observation & 34.52 & 8.66& 15.89 \\
\textbf{EAOG} & Learned active observation & \textbf{45.47} & \textbf{61.92}  & \textbf{34.57} \\
\bottomrule
\end{tabular}
\end{table}

\subsection{Q2: Does Active Observation Improve Recognition?}
\label{sec:exp-active}

We next ask whether actively changing the viewpoint can improve affect
recognition. We evaluate the training-free heuristic baseline, which
uses neither emotion supervision nor recognizer feedback.

\textbf{Broad zero-shot gains.}
On the complete test set, the heuristic baseline improves Macro-F1 for 21 of
24 frozen recognizer configurations (Table \ref{tab:main-results}). Mean Macro-F1
increases from 9.89\% to 11.70\% for the 19 open-source recognizers and from
22.61\% to 24.26\% for the five closed-source recognizers. These gains recover
32.0\% and 20.1\% of the corresponding P-Ref--P-Init observation-quality gaps,
respectively. Thus, active viewpoint acquisition improves affect recognition
even without emotion-specific supervision. Complete per-recognizer results,
Recovery Rate, and path-aware E-SPL results are reported in
Tables \ref{app:tab:main-results-full} and \ref{app:tab:recovery-efficiency-full}.

\textbf{Successful acquisition.}
The complete-set result also reflects acquisition failures. The heuristic
baseline successfully acquires a valid new observation in 1{,}881 of the
3{,}941 test episodes. On this successful-acquisition subset
(Figure \ref{fig:new-video-gains}), the gains over paired P-Init increase to
3.40 and 2.91 percentage points for the open- and closed-source groups,
respectively, with improvements for 20 of the 24 recognizer configurations.
This shows that successfully acquired viewpoints generally provide more useful
affective evidence, while reliable viewpoint acquisition remains an important
bottleneck. Detailed subset results are reported in Table~\ref{app:tab:aobs-provenance}.

\textbf{Learned vs.\ heuristic observation.}
We further compare heuristic and learned active observation using the same
AffectSim-adapted and frozen VideoLLaMA2.1-7B-AV recognizer.
As shown in Table \ref{tab:system_comparison}, P-Init achieves 32.74\% Macro-F1, P-Ref achieves 53.3\%, the heuristic baseline reaches 34.52\%, and EAOG reaches 45.47\%.
This controlled comparison shows that learned active observation provides a
substantial improvement beyond the hand-designed heuristic baseline.
Q3 further isolates the contribution of evidence-aware termination by fixing
the learned navigator and recognizer and varying only the termination strategy.

\begin{table}[t]
\centering
\begin{minipage}[t]{0.49\textwidth}
\centering
\caption{Recognition and acquisition efficiency under different observation-termination strategies. The navigator and recognizer are fixed; only the termination strategy differs.}

\label{tab:eaot-timing-controls}
\small
\setlength{\tabcolsep}{4pt}
\renewcommand{\arraystretch}{1.15}
\begin{tabular*}{\linewidth}{@{\extracolsep{\fill}}lccc@{}}
\toprule
\textbf{Method} & \textbf{F1$\uparrow$} & \textbf{OGR$\uparrow$} & \textbf{E-SPL@3$\uparrow$} \\
\midrule
Horizon mid. (24\,s)   & 43.02 & 50.00 & 32.04 \\
Candidate mid. (30\,s) & 43.94 & 54.47 & 31.89 \\
Final (48\,s)          & 43.93 & 54.43 & 29.79 \\
Uniform random         & 44.27 & 56.08 & 32.18 \\
\midrule
\textbf{EAOG}          & \textbf{45.47} & \textbf{61.92} & \textbf{34.57} \\
\bottomrule
\end{tabular*}
\end{minipage}\hfill%
\begin{minipage}[t]{0.49\textwidth}
\vspace{0pt}
\centering
\caption{Ablation of EAOG training objectives. Each variant removes one training signal while keeping the remaining training and evaluation settings unchanged.}

\label{tab:eaot-loss-ablation}
\small
\setlength{\tabcolsep}{4pt}
\renewcommand{\arraystretch}{1.15}
\begin{tabular*}{\linewidth}{@{\extracolsep{\fill}}lccc@{}}
\toprule
\textbf{Method} & \textbf{F1$\uparrow$} & \textbf{OGR$\uparrow$} & \textbf{E-SPL@3$\uparrow$} \\
\midrule
w/o emotion loss & 42.65 & 48.20 & 31.83 \\
w/o quality loss & 44.65 & 57.93 & 32.52 \\
w/o waiting cost & 42.10 & 45.53 & 31.64 \\
\midrule
\textbf{EAOG}    & \textbf{45.47} & \textbf{61.92} & \textbf{34.57} \\
\bottomrule
\end{tabular*}
\end{minipage}
\end{table}

\subsection{Q3: Can Evidence-Aware Observation Gate Improve Recognition within a Limited Observation Budget?}
\label{sec:exp-learned-system}

\textbf{Adaptive vs.\ non-adaptive termination.}
We compare EAOG with four non-adaptive strategies: terminating at the midpoint
of the full observation horizon (Horizon midpoint, 24\,s), at the midpoint of
the valid candidate interval (Candidate midpoint, 30\,s), at the final
observation (48\,s), or at a uniformly sampled candidate time. These controls
represent simple alternatives that use the same observations but do not adapt
the termination time to the acquired affective evidence.

As shown in Table \ref{tab:eaot-timing-controls}, EAOG achieves the best performance
across all three metrics, with 45.47\% Macro-F1, 61.92\% OGR, and 34.57
E-SPL@3. It outperforms the best fixed-time strategy by 1.53 percentage points
and uniform random termination by 1.20 points in Macro-F1. EAOG terminates at
33.72\,s on average, 14.28\,s earlier than the 48-s endpoint. These results
show that the most useful affective evidence does not necessarily occur at the
end of observation, and that adapting the termination time to accumulated
evidence improves both recognition quality and acquisition efficiency.
Detailed stopping-time and recognition statistics are reported in
Section \ref{app:stopping-controls}.

\textbf{EAOG objective ablation.}
We further ablate the emotion loss, quality loss, and waiting cost while
keeping the remaining training and evaluation settings unchanged
(Table \ref{tab:eaot-loss-ablation}). Removing these components reduces Macro-F1
from 45.47\% to 42.65\%, 44.65\%, and 42.10\%, respectively. In particular,
removing the waiting cost causes the largest degradation, indicating that
explicitly balancing recognition quality against delayed observation is
important for effective termination. The full EAOG also achieves the highest
OGR and E-SPL@3. Additional accuracy, Half-Brier, stopping-time, and
threshold-selection results are provided in Table \ref{app:tab:eaot-method-ablation-full} and Section \ref {app:eaot-threshold-selection}.

\begin{figure}[t]
\centering
\includegraphics[width=\linewidth]{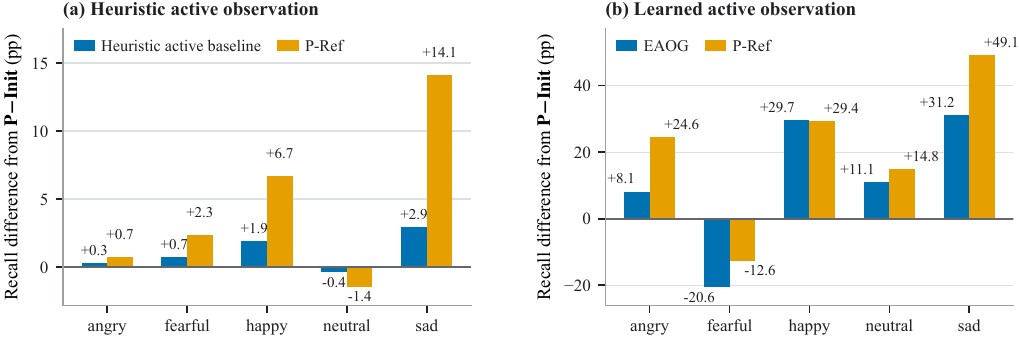}

\caption{
Class-wise recall over the paired P-Init observation.
(a) Heuristic-baseline averaged over all 24 frozen recognizer
configurations.
(b) EAOG under the controlled recognizer setting.
P-Ref is shown as a diagnostic reference.
All values are percentage-point (pp) gains.
}
\label{fig:class-action-diagnostics}
\end{figure}

\subsection{Q4: Where Does Active Observation Help, and What Does It Cost?}
\label{sec:exp-analysis}

\textbf{Class-wise variation.}
As shown in Figure \ref{fig:class-action-diagnostics}, active observation does not
benefit all emotions equally. For the heuristic baseline, the largest average
recall gains occur for sad (+2.9 pp) and happy (+1.9 pp), while neutral
slightly decreases (-0.4 pp). The P-Ref results further show that the
observation-quality gap itself varies substantially across emotion categories.
Under the controlled EAOG setting, the class-wise changes are larger:
recall improves for happy (+29.7 pp), sad (+31.2 pp), neutral (+11.1 pp),
and angry (+8.1 pp), but decreases for fearful (-20.6 pp).
Thus, learned active observation yields substantial overall gains but also
introduces a pronounced class-dependent trade-off. More results are reported in the appendix.

\textbf{Acquisition cost.}
Improved observations are not free to acquire. For the heuristic baseline,
47.73\% of test episodes reach stable handoff, while 52.27\% exhaust the
action budget beforehand. The robot travels 23.23\,m per episode on average,
and 97.4\% of locomotion occurs before handoff. This indicates that most of
the acquisition cost lies in locating and approaching the target rather than
maintaining the final observation. Together with the efficiency results in
Q3, these findings identify viewpoint acquisition itself as a major bottleneck
for active affective perception.
More execution statistics, per-model Recovery Rate and path-aware
E-SPL are reported in Tables \ref{app:tab:recovery-efficiency-full} and \ref{app:tab:embodied-execution}.

\section{Conclusion}

We introduce AffectSim, a controllable interactive 3D benchmark for embodied
affective perception, together with a training-free heuristic baseline and
the learned EAOG method. Our results establish observation quality as a key
factor in affect recognition and demonstrate the value of active observation.
EAOG improves recognition while reducing observation time relative to the full-horizon baseline.
More importantly, AffectSim uses simulation as an experimental instrument
rather than merely a means of scaling data, enabling controlled and repeatable
re-observation that is difficult to realize in the physical world. It provides
a foundation for studying affective perception as an embodied sensing problem.
\textbf{Limitations and Future Directions:}
AffectSim currently focuses on body-motion cues and predefined human behaviors.
Future work can incorporate facial, speech, and other multimodal cues, more
realistic avatars and environments, reactive virtual humans, and real-robot
evaluation toward richer interactive affective intelligence.

\clearpage
\section*{Ethics Statement}

Emotion-perception technologies raise concerns about surveillance,
manipulation, psychological profiling, privacy, and culturally dependent
labels. The motion-capture protocol follows procedures for informed consent,
compensation, data governance, withdrawal, and data use. Released avatars do
not reproduce participants' real facial appearance, reducing direct identity
exposure relative to in-the-wild video. \benchmarkname{} is intended for
research on embodied perception, assistive robotics, and human-centered AI,
not high-stakes decision making or surveillance. Release documentation will
describe label subjectivity and cultural limitations, intended uses, and
procedures for reporting potential misuse.

\section*{AI Use Statement}

Generative AI tools were used to assist with language compression and
\LaTeX{} formatting. The authors remain responsible for verifying all
technical content, citations, claims, and reported results.

\section*{Reproducibility Statement}

Benchmark assets, code, and simulation environments will be released.
Section~\ref{sec:benchmark} describes the episode construction procedure.
Section~\ref{sec:protocol} specifies the paired observation protocol.
The appendix provides complete metric definitions, data split
specifications, execution details, and results for each configuration.

\bibliographystyle{plainnat}
\bibliography{references}

\clearpage
\appendix
\setcounter{table}{0}
\renewcommand{\thetable}{\Alph{table}}
\renewcommand{\theHtable}{appendix.\Alph{table}}
\setcounter{figure}{0}
\renewcommand{\thefigure}{\Alph{figure}}
\renewcommand{\theHfigure}{appendix.\Alph{figure}}

\section*{Appendix Contents}
\noindent\hyperref[app:benchmark-comparison]{Section~\ref*{app:benchmark-comparison}: Detailed Benchmark Comparison}\dotfill\pageref*{app:benchmark-comparison}\par
\noindent\hyperref[app:benchmark-details]{Section~\ref*{app:benchmark-details}: Supplementary Benchmark Construction Details}\dotfill\pageref*{app:benchmark-details}\par
\noindent\hyperref[app:protocol-details]{Section~\ref*{app:protocol-details}: Supplementary Metric Definitions}\dotfill\pageref*{app:protocol-details}\par
\noindent\hyperref[app:training-details]{Section~\ref*{app:training-details}: Training and Implementation Details}\dotfill\pageref*{app:training-details}\par
\noindent\hyperref[app:eaot-implementation]{Section~\ref*{app:eaot-implementation}: EAOG Architecture and Training Details}\dotfill\pageref*{app:eaot-implementation}\par
\noindent\hyperref[app:experimental-results]{Section~\ref*{app:experimental-results}: Additional Experimental Results}\dotfill\pageref*{app:experimental-results}\par
\noindent\hyperref[app:limitations]{Section~\ref*{app:limitations}: Limitations, Ethics, and Future Work}\dotfill\pageref*{app:limitations}\par
\medskip
\clearpage

\section{Detailed Benchmark Comparison}
\label{app:benchmark-comparison}

\subsection{Expanded Related-Work Discussion}
\label{app:expanded-related-work}

\begin{table}[H]
\centering
\setlength{\belowcaptionskip}{\baselineskip}
\caption{
Comparison of representative affective and embodied benchmarks.
\presentmark{} and \absentmark{} indicate whether each capability
is supported in an official evaluation setting.
Active perception refers to acquiring new observations
through agent actions.
}
\label{app:tab:benchmark-comparison}

\begingroup
\renewcommand{\baselinestretch}{1}
\small
\setlength{\tabcolsep}{3pt}
\setlength{\extrarowheight}{0pt}
\renewcommand{\arraystretch}{1.05}
\setlength{\aboverulesep}{0.4ex}
\setlength{\belowrulesep}{0.4ex}

\newcommand{\benchhead}[1]{{\scriptsize\bfseries #1}}

\begin{tabularx}{\textwidth}{@{}
    >{\raggedright\arraybackslash}X
    >{\centering\arraybackslash}p{0.130\textwidth}
    >{\centering\arraybackslash}p{0.095\textwidth}
    >{\centering\arraybackslash}p{0.075\textwidth}
    >{\centering\arraybackslash}p{0.135\textwidth}
    >{\centering\arraybackslash}p{0.115\textwidth}
    >{\centering\arraybackslash}p{0.090\textwidth}
@{}}
\toprule

\multirow{2}{*}{\benchhead{Benchmark}}
& \benchhead{Active}
& \benchhead{Egocentric}
& \benchhead{3D}
& \benchhead{Interactive}
& \benchhead{Controlled}
& \benchhead{Agent}
\\[-2pt]
& \benchhead{perception}
& \benchhead{view}
& \benchhead{humans}
& \benchhead{environment}
& \benchhead{replay}
& \benchhead{control}
\\

\midrule

\rowcolor{benchmarkgroupbg}
\multicolumn{7}{@{}l@{}}{
    \strut\emph{Passive affective benchmarks}
}
\\

DFEW~\citep{jiang2020dfew}
& \absentmark
& \absentmark & \absentmark & \absentmark
& \absentmark & \absentmark
\\

MAFW~\citep{liu2022mafw}
& \absentmark
& \absentmark & \absentmark & \absentmark
& \absentmark & \absentmark
\\

MER2024~\citep{lian2024mer2024}
& \absentmark
& \absentmark & \absentmark & \absentmark
& \absentmark & \absentmark
\\

Emilya~\citep{fourati2014emilya}
& \absentmark
& \absentmark & \absentmark & \absentmark
& \absentmark & \absentmark
\\

KDAE~\citep{zhang2020kinematic}
& \absentmark
& \absentmark & \absentmark & \absentmark
& \absentmark & \absentmark
\\

BoLD~\citep{luo2020arbee}
& \absentmark
& \absentmark & \absentmark & \absentmark
& \absentmark & \absentmark
\\

iMiGUE~\citep{liu2021imigue}
& \absentmark
& \absentmark & \absentmark & \absentmark
& \absentmark & \absentmark
\\

IEMOCAP~\citep{busso2008iemocap}
& \absentmark
& \absentmark & \absentmark & \absentmark
& \absentmark & \absentmark
\\

MELD~\citep{poria2019meld}
& \absentmark
& \absentmark & \absentmark & \absentmark
& \absentmark & \absentmark
\\

EMOTIC~\citep{kosti2017emotic}
& \absentmark
& \absentmark & \absentmark & \absentmark
& \absentmark & \absentmark
\\

MERBench~\citep{lian2026merbench}
& \absentmark
& \absentmark & \absentmark & \absentmark
& \absentmark & \absentmark
\\

DEEMO~\citep{li2025deemo}
& \absentmark
& \absentmark & \absentmark & \absentmark
& \absentmark & \absentmark
\\

MME-Emotion~\citep{zhang2026mmeemotion}
& \absentmark
& \absentmark & \absentmark & \absentmark
& \absentmark & \absentmark
\\

\midrule

\rowcolor{benchmarkgroupbg}
\multicolumn{7}{@{}l@{}}{
    \strut\emph{Egocentric and social-affective benchmarks}
}
\\

Robot-centric Emotion Dataset~\citep{quiroz2022robotcentric}
& \absentmark
& \presentmark & \presentmark & \absentmark
& \absentmark & \absentmark
\\

$E^3$~\citep{lin2024e3}
& \absentmark
& \presentmark & \absentmark & \absentmark
& \absentmark & \absentmark
\\

OSMO~\citep{abdelfattah2026osmo}

& \absentmark
& \presentmark & \absentmark & \absentmark
& \absentmark & \absentmark
\\

RobotEQ~\citep{fang2026roboteq}
& \absentmark
& \presentmark & \absentmark & \absentmark
& \absentmark & \absentmark
\\

EmpathyAgent~\citep{chen2025empathyagent}
& \absentmark
& \presentmark & \presentmark & \absentmark
& \absentmark & \absentmark
\\

\midrule

\rowcolor{black!7}
\textbf{\benchmarkname{} (ours)}
& \presentmark
& \presentmark & \presentmark & \presentmark
& \presentmark & \presentmark
\\

\bottomrule
\end{tabularx}
\endgroup
\end{table}

We retain the fuller discussion here because it explains more precisely which part of the affective-perception pipeline is placed inside the evaluation loop.

\paragraph{Emotion understanding from fixed observations.}
Most affective computing benchmarks evaluate models from observations fixed
before inference. IEMOCAP and MELD study emotion recognition from prerecorded
multimodal conversations
\citep{busso2008iemocap,poria2019meld}; DFEW and MAFW focus on dynamic facial
and multimodal emotion recognition in the wild
\citep{jiang2020dfew,liu2022mafw}; and MER2024 and MERBench provide broader
multimodal evaluation protocols
\citep{lian2024mer2024,lian2026merbench}. Other benchmarks extend the supplied
evidence beyond facial appearance. EMOTIC incorporates scene context, BoLD and
iMiGUE emphasize bodily or non-facial behavior, and structured motion datasets
such as Emilya, KDAE, STEP, and BEAT provide explicit motion-level
representations
\citep{kosti2017emotic,luo2020arbee,liu2021imigue,fourati2014emilya,
zhang2020kinematic,bhattacharya2020step,liu2022beat}. These resources support
facial, audiovisual, contextual, and body-based emotion recognition
\citep{YIN2024110117,Shangguan2025FacialSurvey,Lu_2023,lu2025emotion,
Lu_Chen_Liang_Tan_Zeng_Hu_2025,9607417}.

Recent multimodal foundation models extend affective understanding from
categorical recognition toward explanation and reasoning. Emotion-LLaMA and
EMER model multimodal emotional evidence and its interpretation, while
AffectGPT, EmoBench-M, and MME-Emotion broaden evaluation toward free-form and
reasoning-oriented affective understanding
\citep{cheng2024emotion,lian2023emer,lian2025affectgpt,hu2025emobenchm,
zhang2026mmeemotion}. Despite richer evidence and stronger reasoning, these
settings retain a common assumption: the observation itself is predetermined.
The evaluated model can recognize, fuse, or reason over supplied evidence, but
cannot physically act to improve it or re-observe the same affective behavior
under controlled conditions. \benchmarkname{} studies this upstream
observation process.

\paragraph{Embodied affective perception and social intelligence.}
Recent work has moved affective computing toward robot-centric and egocentric
settings. The Robot-centric Emotion Dataset captures affective behavior during
human--robot interaction, while $E^3$ and OSMO provide first-person
audiovisual observations in everyday or social scenarios
\citep{quiroz2022robotcentric,lin2024e3,abdelfattah2026osmo}. RobotEQ studies
emotional intelligence in embodied scenarios using egocentric observations and
spatial grounding \citep{fang2026roboteq}, while ActFER adaptively selects
facial regions for recognition \citep{liu2026actfer}. These works bring
affective perception closer to the observer, but the physical observation
trajectory remains predetermined: the evaluated system cannot reposition
itself to alter subsequent visual evidence or systematically re-observe the
same affective event.

Beyond perception, systems such as EmpathyAgent integrate affective perception,
memory, reasoning, and action generation for empathetic interaction
\citep{chen2025empathyagent}. Their focus is primarily on how an embodied agent
interprets supplied affective evidence and acts upon it, rather than on how that
evidence should be physically acquired. In contrast, \benchmarkname{} places
the observation process itself inside the evaluation loop. Its executable 3D
episodes support controlled re-observation of the same affective behavior, and
an agent's actions directly change the visual evidence subsequently available
for emotion recognition.

\section{Supplementary Benchmark Construction Details}
\label{app:benchmark-details}

The main construction pipeline, quality-control criteria, and split statistics appear in \cref{sec:benchmark}. This appendix provides motion examples and additional implementation checks.

\begin{figure}[t]
  \centering
  \includegraphics[width=0.88\linewidth]{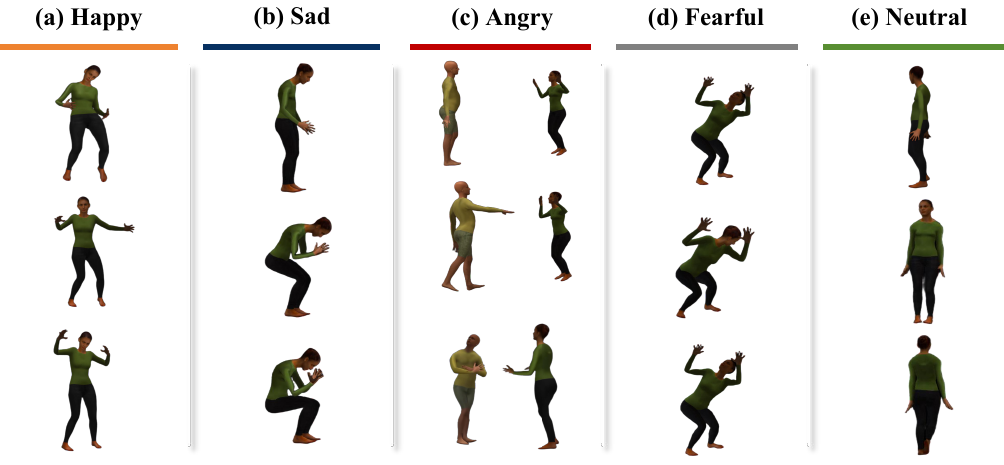}
  \caption{Representative \benchmarkname{} motion assets across five emotion
  categories, covering both single-person and dyadic interactions.}
  \label{app:fig:emotion-examples}
\end{figure}

\subsection{Motion Retargeting Checks}

\begin{table}[t]
\centering
\setlength{\belowcaptionskip}{\baselineskip}
\caption{
Group-disjoint benchmark splits of \benchmarkname{}. All episodes generated
from the same source performance remain in the same split.
}
\label{tab:splits}
\footnotesize
\setlength{\tabcolsep}{3.2pt}
\renewcommand{\arraystretch}{1.05}
\begin{adjustbox}{max width=\textwidth}
\begin{tabular}{lrrrrrrr}
\toprule
\textbf{Split}
& \multicolumn{2}{c}{\textbf{Action-grounded}}
& \multicolumn{2}{c}{\textbf{Observation-controlled}}
& \multicolumn{2}{c}{\textbf{Dyadic pilot}}
& \textbf{Total} \\
\cmidrule(lr){2-3}\cmidrule(lr){4-5}\cmidrule(lr){6-7}
& \textbf{Actors} & \textbf{Episodes}
& \textbf{Motions} & \textbf{Episodes}
& \textbf{Motions} & \textbf{Episodes} & \textbf{Episodes} \\
\midrule
Train      & 8  & 11{,}999 & 669 & 6{,}643 & 34 & 1{,}020 & 19{,}662 \\
Validation & 2  & 3{,}003  & 84  & 831     & 7  & 210      & 4{,}044 \\
Test       & 2  & 2{,}899  & 84  & 832     & 7  & 210      & 3{,}941 \\
\midrule
Total      & 12 & 17{,}901 & 837 & 8{,}306 & 48 & 1{,}440  & 27{,}647 \\
\bottomrule
\end{tabular}
\end{adjustbox}
\end{table}

The resulting motions are screened for skeletal distortions, implausible joint configurations, root-motion discontinuities, foot sliding, and visible body penetration.

\subsection{Episode Initialization and Integrity Checks}

For action-grounded episodes, the robot and target occupy the same accessible
local environment with the target initially visible; for
observation-controlled episodes, $x_0$ is selected according to the assigned
challenge family.

Scene and motion references must be resolvable, episode identifiers unique,
metadata consistent with the release manifest, and robot initial states
navigable.
Motion-specific checks verify support alignment, hand--door geometry,
standing-to-seated transitions, and collision-free walking trajectories.

\subsection{Affective-Semantic Validation after Retargeting}
\label{app:retargeting-validation}

To assess whether motion retargeting preserves the affective semantics of the
source performances, we randomly sampled 500 retargeted motion assets for
manual evaluation. Three reviewers independently inspected the rendered motions
and assigned one of the five AffectSim emotion labels: angry, fearful, happy,
sad, or neutral. The final label for each motion was determined by majority vote.

Compared with the source emotion labels, the retargeted motions achieve an
overall accuracy of 78\% and a Macro-F1 of 77.81\%. Table~\ref{tab:retargeting-validation}
reports the class-wise precision, recall, and F1 scores. These results indicate
that the affective semantics are largely preserved after retargeting.

\begin{table}[t]
\centering
\caption{Class-wise results of manual affective-semantic validation on the
sampled retargeted motions.}
\label{tab:retargeting-validation}
\begin{tabular}{lccc}
\toprule
Emotion & Precision (\%) & Recall (\%) & F1 (\%) \\
\midrule
Angry   & 73.68 & 78.65 & 76.09 \\
Fearful & 87.38 & 79.65 & 83.33 \\
Happy   & 78.43 & 68.97 & 73.39 \\
Sad     & 76.25 & 74.39 & 75.31 \\
Neutral & 74.17 & 89.00 & 80.91 \\
\midrule
Macro avg. & 77.98 & 78.13 & 77.81 \\
\bottomrule
\end{tabular}
\end{table}

\subsection{Simulator Interface and Extensibility}
\label{app:episode-abstraction}

The benchmark can expose egocentric RGB-D observations and
proprioceptive state. Because actions
alter future observations, one episode can support matched initial, reference,
monitor, and actively acquired views. The asset--episode separation also makes
the benchmark extensible: new affective motions, performers, dyadic behaviors,
generated motions, avatars, scenes, and observation challenges can be added
without redefining the episode abstraction or evaluation protocol.

\section{Supplementary Metric Definitions}
\label{app:protocol-details}

The matched observation settings, study-specific budget, and primary metric definitions appear in \cref{sec:protocol}. The following formulas specify the supplementary recovery and efficiency measures.

\paragraph{\textbf{Episode-level recovery}}
For paired P-Init and active predictions, Recovery Rate is
\begin{equation}
  \mathrm{Recovery\ Rate}(\pi)=
  \frac{\sum_i\mathbf{1}[\hat y_i^{\mathrm{P\text{-}Init}}\neq y_i
                          \land\hat y_i^{\pi}=y_i]}
       {\sum_i\mathbf{1}[\hat y_i^{\mathrm{P\text{-}Init}}\neq y_i]}.
  \label{eq:recovery-rate}
\end{equation}

\paragraph{Acquisition efficiency}

At target-distance reference $r$,

\begin{equation}
  \mathrm{E\text{-}SPL}@r
  =\frac{1}{N_r}\sum_{i\in\mathcal{I}_r}
  \mathbf{1}[\hat y_i^{\pi}=y_i]
  \frac{d_i^{\star}(r)}{\max(L_i,d_i^{\star}(r))},
  \qquad r\in\{1,3,5\}\,\mathrm{m}.
  \label{eq:espl}
\end{equation}
Here $\mathcal I_r$ is the set of evaluated episodes whose $r$-conditioned
target region is nonempty and reachable, and $N_r=|\mathcal I_r|$.
The eligible set and denominator are computed separately for each evaluation
cohort. The quantity $d_i^{\star}(r)$ is the shortest collision-free path from the initial
position to a reachable NavMesh point within $0.1$ m of reference distance
$r$, and $L_i$ is the executed path.

Visibility and full-body framing statistics are reported
only when the target identity can be estimated without using prohibited
ground-truth information.

\section{Training and Implementation Details}
\label{app:training-details}

This section specifies the learned navigator's architecture and
behavior-cloning procedure. \textcolor{black}{EAOG} and recognizer details follow in
Appendix~\ref{app:eaot-implementation}.

\subsection{Navigation Architecture}
\label{app:navigation-implementation}

\textcolor{black}{The learned observation policy in Section~\ref{sec:method} adapts
Uni-NaVid~\citep{zhang2025uninavidvideobasedvisionlanguageactionmodel}
with a five-action movement head. The actions are \textsc{Forward},
\textsc{Backward}, \textsc{Left}, \textsc{Right}, and \textsc{Wait};
EAOG makes the separate observation-termination decision.}
At deployment, the navigator uses RGB, the language instruction, and its own
executed actions.
\textcolor{black}{Forward and backward request approximately $0.25$\,m of
translation, left and right request a $15^\circ$ rotation, and wait holds
the pose for one $0.25$-s control step. The fixed instruction is:
\emph{Find the person and maintain a clear, well-framed body view at a
comfortable distance. Keep observing useful body motion; front or side
views are preferred when reachable without losing the person. Move back
when the person approaches, and go around obstacles. Use WAIT when the
current body view is stable and useful.} The student does not receive a
separate text encoder; the instruction is processed by the Uni-NaVid
language-model tokenizer and embedding layer.}

\paragraph{Short-window visual and language encoding.}
Let $n$ index navigation decisions at $4$\,Hz. At each decision, the navigator uniformly samples
up to $F_{\mathrm{nav}}=6$ frames from the preceding $2$\,s and resizes them
to $224\times224$. 
The frozen EVA-CLIP tower produces a class token and a $16\times16$ grid
of 1408-dimensional patch tokens per frame. After removing the class token,
each frame is average-pooled to a $2\times2$ grid for coarse temporal context.
The current frame is additionally pooled to an $8\times8$ grid for finer
spatial detail. A full six-frame window therefore supplies 88 EVA tokens
($6\times4+8\times8$).

A multimodal projector maps the visual tokens from the EVA encoder into the
language-model embedding space. It consists of two linear layers with
dimensions $1408\rightarrow4096\rightarrow4096$, separated by a GELU
activation. The navigation instruction is formatted using the Vicuna
conversation template, tokenized by the language model's tokenizer, and
converted into text embeddings by its input embedding layer. The projected
visual tokens replace the image placeholder in the token sequence, after
which the visual and text embeddings are jointly processed by the LLaMA
backbone. The hidden state of the final valid multimodal token forms the fused representation
$\mathbf{h}_n^{\mathrm{VLA}}\in\mathbb{R}^{4096}$.

\paragraph{Parameter-efficient adapters.}
A residual adapter is inserted into each of the last four LLaMA blocks.
For the hidden state $\mathbf H^{(l)}$ of block $l$, it computes
\begin{equation}
    \widetilde{\mathbf H}^{(l)}
    =\mathbf H^{(l)}+
    \operatorname{Dropout}\!\left(
      \mathbf W_{\mathrm{up}}^{(l)}
      \operatorname{GELU}\!\left(
        \mathbf W_{\mathrm{down}}^{(l)}
        \operatorname{LN}(\mathbf H^{(l)})\right)\right),
    \label{eq:navigation_adapter}
\end{equation}
where the down and up projections have dimensions $64\times4096$ and
$4096\times64$, respectively. Zero initialization of the up projection
preserves the pretrained representation when the adapter is first inserted.

\paragraph{Action history and movement prediction.}
The policy conditions on the two most recently executed actions and the
run length $r_n$ of the current repeated action. A shared embedding table
with six entries and width 32 represents \textsc{Start} and the five actions.
A second table with five entries and width 16 represents the run-length
bins $\{0,1,2\text{--}3,4\text{--}7,8+\}$. Their concatenation is
\begin{equation}
    \mathbf x_n=
      [\mathbf h_n^{\mathrm{VLA}};
       \mathbf e_a(a_{n-1});\mathbf e_a(a_{n-2});\mathbf e_r(r_n)]
      \in\mathbb R^{4176}.
    \label{eq:navigation_policy_state}
\end{equation}
The movement head maps this state to five action logits:
\begin{equation}
    \mathbf z_n^{\mathrm{nav}}
      =\mathbf W_2\operatorname{Dropout}\!\left(
        \operatorname{GELU}\!\left(
          \mathbf W_1\operatorname{LN}(\mathbf x_n)\right)\right),
    \label{eq:navigation_head}
\end{equation}
with layer widths $4176\rightarrow1024\rightarrow5$.
Greedy inference executes $\arg\max_a z_{n,a}^{\mathrm{nav}}$.
Only actions acknowledged by the simulator enter the executed-action history.

In the final training pass, the EVA tower and original
7B language model weights are frozen. Trainable components are the four adapters, action history embeddings, movement head, and final RMSNorm. After training, these components are also frozen for \textcolor{black}{EAOG} training and evaluation.

\subsection{{Supervision Signal and Behavior Cloning}}
\label{app:navigation-training}

\paragraph{Privileged observation expert.}
During supervision generation, the expert accesses target world position
and motion trend, scene NavMesh, executable paths, person--robot distance,
and rendered-body visibility. It balances navigability, standoff distance,
full-body visibility, and stable observation. Its actions include backing
away from an approaching person, turning to reacquire a lost target,
avoiding non-executable translations, and waiting to preserve a useful view.
These privileged variables are excluded from student inputs.
\textcolor{black}{The expert also checks whether a proposed translation is
executable and can leave a state unlabeled when its action supervision is
unreliable. The simulator restores the episode's scene instance, matching
NavMesh, robot and person initialization, and motion phase before rollout;
replayed human motion continues cyclically. At evaluation, none of the
expert's geometry, target pose, body visibility, future motion, or emotion
label is supplied to the student.}

\paragraph{Initial behavior cloning.}
\textcolor{black}{The expert corpus covers training episodes. An event-balanced view retains
480{,}000 labeled state--action pairs, selecting 12--40 expert decisions per
episode while maintaining coverage of this corpus. Each pair contains the
student-observable state and its expert action label; these are decision
samples, not additional dataset episodes. Selection preserves initial states,
action transitions,
backward motion, person approach, visibility changes, target loss and
reacquisition, informative-view neighborhoods, and motion-loop boundaries.
The final adaptation uses two epochs with effective batch size eight;
the frozen visual tower, multimodal projector, and original language-model
weights are not updated.}

\section{{EAOG} Architecture and Training Details}
\label{app:eaot-implementation}

During \textcolor{black}{EAOG} training, the navigator, EVA visual tower, and affect
recognizer remain frozen. The \textcolor{black}{quality} and emotion heads supervise the
evidence states, and the \textcolor{black}{termination} head selects the window submitted
for recognition.

\subsection{Candidate Windows and Observation Budget}
\label{app:eaot-windows}

The navigator acts at $4$\,Hz. \textcolor{black}{EAOG} first evaluates at $12$\,s and then once
per second through the $48$\,s deadline.
Each window $\mathcal W_t$ contains $F=32$ deterministically sampled frames
from the most recent 48 RGB observations, corresponding to $12$\,s of
observation. The affect recognizer receives only
the selected window, not the recurrent state or the complete observation history.

\subsection{Visual Tokens and Emotion Evidence}
\label{app:eaot-architecture}

\paragraph{Visual token construction.}
\textcolor{black}{EAOG} uses the EVA-CLIP visual tower of Uni-NaVid, bypassing the multimodal
projector and language model. Removing the class token leaves a
$16\times16$ grid of 1408-dimensional patch features for each frame.
Average pooling reduces this to a $2\times2$ spatial grid. The 32 frames
are divided into six contiguous, nearly equal-sized temporal bins.
For bin $\mathcal B_u$ and spatial cell $v$, the pooled feature is
\begin{equation}
    \mathbf v_{t,u,v}
    =\frac{1}{|\mathcal B_u|}\sum_{f\in\mathcal B_u}
      \widetilde{\mathbf P}_{t,f,v},
    \qquad u=1,\ldots,6,\quad v=1,\ldots,4,
    \label{app:eq:eaot-pooling}
\end{equation}
where $\widetilde{\mathbf P}$ denotes the spatially pooled EVA features.
The resulting 24 tokens retain coarse temporal and spatial structure;
averaging does not preserve frame order within a temporal bin.

\paragraph{Token encoding and emotion queries.}
Each token is normalized, projected to $d=128$, and given learned temporal
and spatial embeddings, $\mathbf a_u$ and $\mathbf b_v$:
\begin{equation}
    \mathbf x_{t,u,v}
    =\mathbf W_p\operatorname{LN}(\mathbf v_{t,u,v})
      +\mathbf a_u+\mathbf b_v.
    \label{eq:eaot_projection}
\end{equation}
A single-layer, four-head Transformer encoder maps these tokens to
$\mathbf H_t\in\mathbb R^{24\times d}$.
Five learned query vectors form $\mathbf Q\in\mathbb R^{5\times d}$.
Their fixed output indices correspond to angry, fearful, happy, neutral,
and sad through the auxiliary emotion loss. They extract evidence as
\begin{align}
    \mathbf Z_t^{(1)}
      &=\operatorname{LN}\bigl(\mathbf Q+
        \operatorname{MHA}(\mathbf Q,\mathbf H_t,\mathbf H_t)\bigr),\\
    \mathbf Z_t
      &=\operatorname{LN}\bigl(\mathbf Z_t^{(1)}+
        \operatorname{FFN}(\mathbf Z_t^{(1)})\bigr).
    \label{eq:emotion_queries}
\end{align}
Each row $\mathbf z_{t,c}$ is a class-conditioned feature vector used by
the recurrent memory and auxiliary emotion readout.
\textcolor{black}{The query vectors are randomly initialized trainable
parameters. At every candidate time, the same five vectors query that
window's encoded visual tokens; the resulting $\mathbf z_{t,c}$ changes with
the current view. They are not updated by the frozen recognizer's predictions
at test time. The auxiliary emotion loss associates their fixed indices with
the five labels, while cross-attention lets them learn distinct
class-conditioned visual summaries. The pooled tokens preserve coarse
within-window temporal and spatial structure, but the model does not retain
all 32 frame-level token grids.}

\paragraph{Recurrent memory and output heads.}
The five query streams share a GRU with input and hidden dimension 128,
but maintain separate recurrent states:
\begin{equation}
    \mathbf e_{t,c}=\operatorname{GRU}
      (\mathbf z_{t,c},\mathbf e_{t-1,c}),
    \qquad c=1,\ldots,5.
    \label{eq:eaot_gru}
\end{equation}
Here $t-1$ denotes the previous one-second candidate, and the recurrent
history is confined to the current episode. The concatenation
$\overline{\mathbf e}_t=[\mathbf e_{t,1};\ldots;\mathbf e_{t,5}]$
has dimension 640. Its readouts are
\begin{equation}
    \ell^{\mathrm{emo}}_{t,c}=\mathbf w_e^\top\mathbf e_{t,c}+b_e,
    \qquad
    \widehat r_t=f_{\mathrm{quality}}(\overline{\mathbf e}_t),
    \qquad
    h_t=\sigma\bigl({\color{black}g_{\mathrm{term}}}(\overline{\mathbf e}_t)\bigr).
    \label{eq:eaot_outputs}
\end{equation}
The quality and \textcolor{black}{termination} heads are separate multilayer perceptrons.
The quality output $\widehat r_t$ predicts standardized recognition risk;
it is not an input to the \textcolor{black}{termination} head. Both auxiliary heads influence the
shared representation during training and are unnecessary for the deployment
decision. \textcolor{black}{EAOG} contains approximately $0.72$M trainable parameters.
\textcolor{black}{All five recurrent states start at zero at the beginning of
each episode. The streams share GRU weights but have separate hidden states,
so a new candidate can change the evidence for each emotion differently.
The termination head sees their concatenation. Its $h_t$ is a conditional
termination probability reflecting whether further observation remains
worthwhile, rather than emotion confidence; the quality and emotion heads
are used only for training supervision. The complete EAOG module has
716{,}163 trainable parameters.}

\subsection{Recognizer Scores and Training Objectives}
\label{app:eaot-supervision}

\paragraph{Candidate-label scoring.}
The frozen recognizer described in Sec.~\ref{sec:exp-setup}
scores each of the five candidate answers under a fixed affect prompt $q$.
\textcolor{black}{It is the visual-only, 32-frame VideoLLaMA2.1-7B-AV
checkpoint at epoch 1, global step 7373, obtained by QLoRA adaptation on the three passive training views
P-Init, P-Ref, and P-Mon (19{,}662 training episodes times three views).
The checkpoint is frozen before query-module supervision, validation, and
test evaluation. Its five answer candidates are \emph{Angry},
\emph{Fearful}, \emph{Happy}, \emph{Neutral}, and \emph{Sad}.}
Let $(w_{c,1},\ldots,w_{c,L_c})$ be the token sequence for answer $a_c$,
including its end-of-sequence token. Its score and normalized probability are
\begin{equation}
    s_{t,c}=\sum_{k=1}^{L_c}\log P_\phi
      (w_{c,k}\mid\mathcal W_t,q,w_{c,<k}),
    \qquad
    p_{t,c}=\frac{\exp(s_{t,c}/T_{\mathrm{cal}})}
      {\sum_{c'=1}^{5}\exp(s_{t,c'}/T_{\mathrm{cal}})}.
    \label{eq:vlm_posterior}
\end{equation}
\textcolor{black}{The same sum-log-probability scores are used for offline
supervision and the selected test window. In this equation,
$T_{\mathrm{cal}}>0$ calibrates probabilities for the offline risk expert;
the direct recognition interface uses $T=1$ for its raw five-candidate
probabilities. Both have the same argmax. Before the final query, neither
form of recognizer probability is available to EAOG.}
\textcolor{black}{Scores sum token log probabilities, including the
end-of-sequence token, without length normalization or free-form generation.
The probability-risk expert uses a scalar temperature fitted on training
candidates ($T_{\mathrm{cal}}=1.4424$ in the supplied training record);
validation and test labels do not refit it. A positive temperature leaves
the predicted class unchanged but affects Half-Brier risk and the soft
stopping target. The prompt is fixed by the dataset's target-person
protocol: a single-person prompt asks for the visible person's dominant
emotion, while dyadic prompts specify the yellow-shirted man,
green-shirted woman, or either eligible person. No prompt is rewritten
using the ground-truth emotion or an intermediate prediction.}
\begingroup\color{black}
\paragraph{Fixed recognizer prompts.}
The answer labels and end-of-sequence scoring rule above are shared by four
prompt variants selected from the episode's target-person metadata:
\begin{description}
\item[Single person:] \emph{Watch the video and classify the dominant
emotional state of the visible person. Choose exactly one label: Angry,
Fearful, Happy, Neutral, Sad. Return only the label.}
\item[Yellow man:] \emph{Classify the body emotion of the specified target
using only the video. The target is the man wearing a yellow top. Choose
exactly one label: Angry, Fearful, Happy, Neutral, Sad. Return only the label.}
\item[Green woman:] \emph{Classify the body emotion of the specified target
using only the video. The target is the woman wearing a green top. Choose
exactly one label: Angry, Fearful, Happy, Neutral, Sad. Return only the label.}
\item[Either actor:] \emph{Classify the body emotion of the specified target
using only the video. Either of the two visible people is an eligible target.
Choose exactly one label: Angry, Fearful, Happy, Neutral, Sad. Return only
the label.}
\end{description}
\endgroup

\paragraph{Recognition loss and soft stopping targets.}
For the one-hot episode label $\mathbf y$, the Half-Brier score is
\begin{equation}
    R_t=\frac12\sum_{c=1}^{5}(p_{t,c}-y_c)^2.
    \label{eq:half_brier}
\end{equation}
This is a probability-prediction loss rather than a classification error
indicator. In Eq.~\ref{eq:time_loss}, time is measured in seconds,
$\lambda=0.002$, and $\tau=0.005$.
The expert assigns more mass to windows with smaller recognition loss
and waiting cost. It uses every candidate on the complete training rollout,
including future windows relative to an intermediate decision. This
privileged expert supplies training targets, while each student prediction
uses only the visual prefix available at that candidate time.
\textcolor{black}{The supplied training record contains 533{,}085 legal candidate windows.
These counts refer to rollouts and candidate windows, respectively, not
additional benchmark episodes. Candidate sequences can be shorter than the
48-s maximum; their padding is masked, and recurrent states are updated
only over valid candidates.}

\paragraph{First-query distribution.}
\textcolor{black}{For each rollout, let $t_{\mathrm{end}}=\max\mathcal T$ be
its last legal candidate (48\,s for a complete online episode). The hazards
define a distribution over the first query, with remaining probability mass
assigned to this forced final candidate:}
\begin{equation}
    \color{black}
    \rho_t=
    \begin{cases}
      h_t\displaystyle\prod_{j\in\mathcal T:j<t}(1-h_j),&t<t_{\mathrm{end}},\\[4pt]
      \displaystyle\prod_{j\in\mathcal T:j<t_{\mathrm{end}}}(1-h_j),
        &t=t_{\mathrm{end}}.
    \end{cases}
    \label{eq:first_stop_distribution}
\end{equation}
We minimize the cross-entropy from the expert distribution,
\begin{equation}
    \color{black}
    \mathcal L_{\mathrm{term}}
    =-\sum_{t\in\mathcal T}d_t\log\rho_t.
    \label{eq:hazard_loss}
\end{equation}
The terminal hazard is not needed because stopping at the deadline is forced.
Equivalently, this objective is a weighted binary cross-entropy over
nonterminal hazards, with target
$h_t^*=d_t/S_t$ and weight $S_t=\sum_{j\in\mathcal T:j\geq t}d_j$.
Thus the sequence objective couples the per-time targets through the
remaining expert probability mass.

\paragraph{Auxiliary supervision.}
\textcolor{black}{Training-candidate statistics standardize the risk as
$\widetilde R_t=(R_t-\mu_R)/\sigma_R$; the supplied training record gives
$\mu_R=0.328612$ and $\sigma_R=0.167778$. The two auxiliary losses are}
\begin{align}
    \mathcal L_{\mathrm{quality}}
      &=\frac{1}{|\mathcal T|}\sum_{t\in\mathcal T}
        \operatorname{SmoothL1}(\widehat r_t,\widetilde R_t),\\
    \mathcal L_{\mathrm{emotion}}
      &=-\frac{{\color{black}0.5}}{|\mathcal T|}\sum_{t\in\mathcal T}\sum_{c=1}^{5}
        y_c\log\operatorname{softmax}(\boldsymbol\ell_t^{\mathrm{emo}})_c.
    \label{app:eq:eaot-auxiliary-losses}
\end{align}
The complete objective is

\begin{equation}
    \color{black}
    \mathcal L_{\mathrm{EAOG}}
      =\mathcal L_{\mathrm{term}}
       +\mathcal L_{\mathrm{quality}}
       +\mathcal L_{\mathrm{emotion}}.
    \label{eq:eaot_total_loss}
\end{equation}

Risk regression encourages features predictive of the recognizer's loss on
the current window. Emotion classification associates the five evidence streams
with the episode label.
\textcolor{black}{The quality target is the risk of the \emph{current}
candidate, not the best risk observed later. SmoothL1 regression teaches the
shared representation to distinguish usable from weak evidence. The
emotion loss gives the five randomly initialized query slots their fixed
class indexing; the shared readout is discarded for final recognition.
Neither auxiliary head is an explicit input to the termination head. The stopping
loss is the only objective that directly trains the first-query distribution.}
\textcolor{black}{EAOG is optimized with AdamW for 50 epochs using batches of
eight episode sequences and gradient clipping at 1.0; candidate padding is
masked. The supplied run uses seeds 20260918 and selects the stated training epoch without using test outcomes.}

\subsection{Online Execution and Query Accounting}
\label{app:eaot-inference}

During the first $12$\,s the frozen navigator acquires observations.
At each subsequent candidate time, \textcolor{black}{EAOG} constructs the current window,
updates its recurrent states, and evaluates $h_t$. If $h_t\geq\theta$,
observation ends; otherwise navigation continues until the next candidate.
At $48$\,s the external controller ends observation regardless of the
predicted hazard.

\subsection{Validation Selection of the Query Threshold}
\label{app:eaot-threshold-selection}

For each trained \textcolor{black}{EAOG} variant, we sweep the stopping threshold on validation
trajectories and select the value with the highest validation Macro-F1 among
those whose mean query time is at most 36\,s, or 75\% of the 48-s observation
horizon. The forced 48-s queries are included when computing this mean. The
selected threshold is then fixed for every test episode of that variant;
test predictions are not used for threshold selection. The resulting values
are $\theta=0.30$ for \textcolor{black}{EAOG}.

The \textcolor{black}{full}
\textcolor{black}{EAOG} test result of 45.47\% Macro-F1 at a 33.72-s mean query time in
Table~\ref{app:tab:eaot-method-ablation-full} therefore uses $\theta=0.30$.
Because each ablation has its own validation-selected threshold, its test
recognition score and mean query time should be interpreted together.

The reported recognition call scores all five candidate answers for
the selected window. Online computation also includes the navigation
model, EVA encoding, and \textcolor{black}{EAOG} updates.

\section{{Additional Experimental Results}}
\label{app:experimental-results}

\begin{table}[t]
  \centering
  \setlength{\belowcaptionskip}{\baselineskip}
  \caption{
    Macro-F1 (\%) on the complete 3{,}941-episode test split.
    Within each row, all observation conditions use the same recognizer and
    episode set. P-Init is the embodied initial observation, P-Ref is the
    privileged reference observation, and P-Mon is the fixed room-monitor
    observation. \textcolor{black}{Heur. denotes the training-free heuristic baseline, with
    paired P-Init fallback when no valid new observation is acquired.}
    $\Delta_{\mathrm{gain}}$ is the displayed \textcolor{black}{Heur.} value minus P-Init.
  }
  \label{app:tab:main-results-full}

  \begingroup
  \setlength{\tabcolsep}{5pt}
  \renewcommand{\arraystretch}{1.03}
  \resizebox{\textwidth}{!}{
  \begin{tabular}{
    lc
    >{\color{black}}c
    >{\color{black}}c
    c
    >{\color{black!75!black}}c
    >{\color{black!75!black}}c
  }
    \toprule

\makecell[l]{\bfseries Perception model}
& \makecell{\bfseries Parameters}
& \makecell{\bfseries\color{black}P-Mon\\[-2pt]
  \scriptsize Monitor view}
& \makecell{\bfseries\color{black}P-Ref\\[-2pt]
  \scriptsize Reference view}
& \makecell{\bfseries P-Init\\[-2pt]
  \scriptsize Initial view}
& \makecell{\bfseries\color{black} \textcolor{black}{Heur.}\\[-2pt]
  \scriptsize \textcolor{black}{Heuristic baseline}}
& \makecell{\bfseries\boldmath\color{black}
  $\Delta_{\mathrm{gain}}$ $\uparrow$\\[-2pt]
  \scriptsize \textcolor{black}{Heur.} $-$ P-Init} \\

\midrule

    \rowcolor{benchmarkgroupbg}
    \multicolumn{7}{l}{
      \strut\textbf{\emph{Open-source general-purpose}} models
    } \\

    InternVL2.5
    \citep{chen2025expandingperformanceboundariesopensource}
    & 4B & 9.92 & 20.60 & 10.04 & \textbf{12.81} & +2.77 \\

    InternVL2.5
    & 8B & 11.36 & 21.90 & 10.88 & \textbf{13.51} & +2.63 \\

    InternVL2.5 INT8
    & 38B & 11.39 & 19.97 & 10.74 & \textbf{13.22} & +2.48 \\

    InternVL2.5
    & 78B & 20.70 & 30.02 & 17.24 & \textbf{18.52} & +1.28 \\

    \addlinespace[2pt]

    Qwen2.5-VL
    \citep{bai2025qwen25vltechnicalreport}
    & 7B & 5.99 & 6.12 & 6.09 & \textbf{6.11} & +0.02 \\

    Qwen2.5-VL INT8
    & 32B & 5.92 & 6.21 & 6.12 & 6.06 & $-$0.06 \\

    Qwen2.5-VL BF16
    & 72B & 6.63 & 7.57 & 6.53 & \textbf{7.63} & +1.10 \\

    Qwen2.5-Omni visual-only
    \citep{xu2025qwen25omnitechnicalreport}
    & 7B & 7.78 & 12.09 & 7.78 & \textbf{9.45} & +1.67 \\

    Qwen3-VL-Instruct
    \citep{bai2025qwen3vltechnicalreport}
    & 8B & 11.79 & 15.54 & 10.12 & \textbf{13.19} & +3.07 \\

    Qwen3-VL-Thinking
    & 8B & 8.68 & 11.15 & 7.93 & \textbf{9.59} & +1.66 \\

    \addlinespace[2pt]

    VideoLLaMA2
    \citep{cheng2024videollama}
    & 7B & 7.47 & 12.69 & 7.08 & \textbf{10.23} & +3.15 \\

    VideoLLaMA2
    & 72B & 10.25 & 19.05 & 9.80 & \textbf{14.08} & +4.28 \\

    VideoLLaMA2.1-16F
    & 7B & 7.97 & 16.25 & 8.73 & \textbf{11.20} & +2.47 \\

    VideoLLaMA2.1-AV visual-only
    & 7B & 20.27 & 27.65 & 18.67 & \textbf{20.89} & +2.22 \\

    \addlinespace[2pt]

    InternVideo2-Chat
    \citep{wang2024internvideo2}
    & 8B & 7.04 & 9.98 & 7.04 & \textbf{8.07} & +1.03 \\

    LongVA-DPO
    \citep{zhang2024long}
    & 7B & 8.95 & 13.57 & 7.55 & \textbf{10.80} & +3.25 \\

    MiniCPM-V2.6
    \citep{yao2024minicpm}
    & 8B & 6.34 & 9.89 & 6.55 & \textbf{6.79} & +0.24 \\

    \addlinespace[2pt]
    \rowcolor{benchmarkgroupbg}
    \multicolumn{7}{l}{
      \strut\textbf{\emph{Open-source emotion-recognition}} models
    } \\

    AffectGPT-frame
    \citep{lian2025affectgpt}
    & 7B & 11.63 & 10.11 & 11.79 & 11.13 & $-$0.66 \\

    Emotion-LLaMA
    \citep{cheng2024emotion}
    & 7B & 15.89 & 25.29 & 17.16 & \textbf{19.04} & +1.88 \\

    Open-source mean (19)
    & -- & 10.31 & 15.56 & 9.89
    & \textbf{11.70} & +1.81 \\

    \midrule
    
    \rowcolor{benchmarkgroupbg}
    \multicolumn{7}{l}{
      \strut\textbf{\emph{Closed-source general-purpose}} models
    } \\
    
    Gemini-3.0-Flash
    & -- & 25.11 & 28.33 & 24.77 & 24.62 & $-$0.15 \\

    Gemini-3.0-Pro
    & -- & 25.21 & 30.21 & 22.66 & \textbf{25.42} & +2.76 \\

    GPT-5.2
    & -- & 21.08 & 28.20 & 20.89 & \textbf{21.00} & +0.11 \\

    GPT-5.6-sol
    & -- & 21.53 & 31.48 & 19.29 & \textbf{21.87} & +2.58 \\

    Claude Opus 5
    & -- & 29.27 & 35.68 & 25.46 & \textbf{28.37} & +2.91 \\

    Closed-source mean (5)
    & -- & 24.44 & 30.78 & 22.61
    & \textbf{24.26} & +1.65 \\

    \bottomrule
  \end{tabular}}
  \endgroup
\end{table}

\subsection{{Inference and Execution Settings}}
\label{app:exp-setup-details}
\paragraph{Perception models and inference}
The zero-shot study evaluates 19 open-source and five closed-source
model configurations on the test split. A configuration denotes a
specific model, scale, quantization setting, and inference interface.
These recognizers are used without \benchmarkname{} adaptation.
The learned-system study uses a separately adapted and frozen
visual-only VideoLLaMA2.1-7B-AV recognizer.

Most open-source general-purpose vision-language models use eight uniformly sampled frames resized to $224\times224$. VideoLLaMA2.1-16F follows the same frame budget, whereas Emotion-LLaMA uses its native single-frame visual input.
Closed-source models receive either the full source-duration video or eight
uniformly sampled frames according to their supported interface. For each
model configuration, the inference interface remains unchanged across all
observation conditions.

\paragraph{\textcolor{black}{Zero-shot heuristic baseline}}
\textcolor{black}{The training-free heuristic baseline first uses a frozen
Uni-NaVid navigation planner~\citep{zhang2025uninavidvideobasedvisionlanguageactionmodel}
to search for the target person. After handoff, a rule-based controller uses
person location and depth to keep the target visible and adjust the observation
viewpoint. The baseline uses neither emotion supervision nor recognizer
feedback, and has no access to oracle trajectories or simulator masks. In
the 24-recognizer zero-shot evaluation, if no valid new observation is acquired,
the recognizer receives the paired P-Init observation. Robot actions run at $4\,\mathrm{Hz}$ with $0.25\,\mathrm{m}$
translations and $15^\circ$ rotations, for at most 192 primitive actions.
After handoff, the acquired recognition clip spans twice the source motion
duration, matching P-Init and P-Ref.}

\paragraph{\textcolor{black}{EAOG}}
\textcolor{black}{The frozen learned observation policy runs at $4\,\mathrm{Hz}$. At a legal query time, the adapted recognizer receives 32 frames sampled from the preceding 12\,s. Observation ends after one query or at the 48-s deadline. Architecture, expert construction, and optimization details are in Appendices~\ref{app:training-details} and~\ref{app:eaot-implementation}.}

\begin{table}[t]
\color{black}
\centering
\setlength{\belowcaptionskip}{\baselineskip}
\caption{\textcolor{black}{Recognition and mean observation-termination time for P-Init, the heuristic baseline, and EAOG using the AffectSim-adapted, frozen VideoLLaMA2.1-7B-AV recognizer. Each EAOG variant uses a validation-selected threshold (Appendix~\ref{app:eaot-threshold-selection}); the full EAOG retains all losses. A dash denotes an unreported metric.}}
\label{app:tab:eaot-method-ablation-full}
\footnotesize
\setlength{\tabcolsep}{4pt}
\renewcommand{\arraystretch}{1.12}

\begin{tabular*}{\textwidth}{@{\extracolsep{\fill}}lrrrr@{}}
\toprule
\textbf{Method / operating point}
& \textbf{F1 (\%) $\uparrow$}
& \textbf{Acc. (\%) $\uparrow$}
& \textbf{Half-Brier $\downarrow$}
& \textbf{Time (s) $\downarrow$} \\
\midrule
\rowcolor{benchmarkgroupbg}
\multicolumn{5}{@{}l}{\textbf{\emph{Fixed initial view}}} \\
P-Init
& 32.74 & 33.95 & \pendingresult & \pendingresult \\
\midrule
\rowcolor{benchmarkgroupbg}
\multicolumn{5}{@{}l}{\textbf{\emph{Heuristic active observation}}} \\
Heuristic baseline
& 34.52 & 38.24 & 0.3874& \pendingresult \\
\midrule
\rowcolor{benchmarkgroupbg}
\multicolumn{5}{@{}l}{\textbf{\emph{Learned active observation with \textcolor{black}{EAOG}}}} \\
Without emotion loss 
& 42.65 & 43.96 & 0.3789 & 32.44 \\
Without quality loss 
& 44.65 & 46.04 & 0.3712 & 35.28 \\
Without waiting cost 
& 42.10 & 43.33 & 0.3842 & 30.70 \\
\addlinespace[2pt]
\textcolor{black}{EAOG}
& 45.47 & 46.25 & 0.3789 & 33.72 \\

\bottomrule
\end{tabular*}
\end{table}

\subsection{{Full Zero-Shot Observation Results}}
\label{app:full-zero-shot-results}
\textcolor{black}{Table~\ref{app:tab:main-results-full} gives the 24 configuration-level scores summarized in main-paper Table~\ref{tab:main-results}. The group means average configurations equally.}

\paragraph{\textbf{\textcolor{black}{Observation-quality gap}}}
On average, P-Ref yields substantially higher Macro-F1 than P-Init. Across the 19 open-source configurations, mean Macro-F1 rises from
9.89\% under P-Init to 15.56\% under P-Ref. For the five closed-source models,
it rises from 22.61\% to 30.78\%. P-Ref achieves the highest score for
18 of 19 open-source configurations and 23 of 24 configurations overall.

\paragraph{\textbf{\textcolor{black}{External monitoring}}}
P-Mon provides a fixed room-level external view as an additional diagnostic.
For open-source models, P-Mon reaches 10.31\% Macro-F1, only slightly above
the 9.89\% obtained by P-Init and substantially below the 15.56\% of P-Ref.
Thus, simply providing an external scene-level camera does not reproduce the
benefit of a deliberately informative observation trajectory. Observation
quality depends jointly on target visibility, framing, orientation, distance,
and the discriminative motion cues exposed to the recognizer.

\subsection{{New-Observation Subset}}
\label{app:new-observation-results}
\textcolor{black}{Heuristic baseline acquired a valid new post-handoff clip in 1{,}881 of 3{,}941 test episodes. The complete-protocol result also includes 2{,}060 episodes with P-Init fallback.}

\begin{table}[t]
\centering
\setlength{\belowcaptionskip}{\baselineskip}
\caption{
Macro-F1 (\%) on the complete test set and the 1{,}881-episode
new-observation subset defined by successful post-handoff acquisition.
\textcolor{black}{In the remaining 2{,}060 episodes, heuristic baseline uses the paired P-Init input.
Gain denotes Heur.$-$P-Init.}
}
\label{app:tab:aobs-provenance}

\footnotesize
\setlength{\tabcolsep}{5pt}
\renewcommand{\arraystretch}{1.06}

\begin{tabular}{cccccc}
\toprule
\textbf{Model group}
& \textbf{Evaluation subset}
& \textbf{Episodes}
& \textbf{P-Init}
& \makecell{\textbf{\textcolor{black}{Heur.}}\\[-2pt]\scriptsize Zero-shot policy}
& \textbf{\textcolor{black}{Heur.}$-$P-Init} \\
\midrule

\multirow{2}{*}{Open-source (19)}
& Complete test set
& 3{,}941
& 9.89
& \textbf{11.70}
& \textbf{+1.81} \\

& New-observation subset
& 1{,}881
& 10.95
& \textbf{14.35}
& \textbf{+3.40} \\

\addlinespace[2pt]

\multirow{2}{*}{Closed-source (5)}
& Complete test set
& 3{,}941
& 22.61
& \textbf{24.26}
& \textbf{+1.65} \\

& New-observation subset
& 1{,}881
& 25.38
& \textbf{28.29}
& \textbf{+2.91} \\

\bottomrule
\end{tabular}
\end{table}

\paragraph{\textbf{\textcolor{black}{Observation gap recovery}}}
Using Observation Gap Recovery (OGR) computed before score rounding, \textcolor{black}{heuristic baseline} recovers 32.0\%
of the mean P-Ref--P-Init gap for open-source models and 20.1\% for
closed-source models. Active movement therefore closes a measurable, although
still incomplete, portion of the gap to the privileged reference observation.
\textcolor{black}{On the new-observation subset, heuristic baseline outperforms P-Init for 20 of 24 configurations (main-paper Figure~\ref{fig:new-video-gains}).}

\begin{table}[t]
  \centering
  \setlength{\belowcaptionskip}{\baselineskip}
\caption{
Recovery Rate and E-SPL@$r$ (\%) on the complete test split.
Recovery Rate measures the fraction of incorrect P-Init predictions corrected
under \textcolor{black}{heuristic baseline}. E-SPL weights final \textcolor{black}{heuristic baseline} recognition correctness by path
efficiency relative to a distance-conditioned target region at
$r\in\{1,3,5\}$~m. Empty or unreachable target regions are excluded, yielding
3,811, 3,887, and 3,859 valid episodes at 1, 3, and 5~m, respectively.
}
  \label{app:tab:recovery-efficiency-full}

  \footnotesize
  \setlength{\tabcolsep}{7pt}
  \renewcommand{\arraystretch}{1.03}

  \begin{adjustbox}{max width=\textwidth}
  \begin{tabular}{lcrrrr}
    \toprule
    \makecell[l]{\bfseries Perception model}
    & \makecell{\bfseries Parameters}
    & \makecell{\bfseries Recovery\\\bfseries Rate}
    & \makecell{\bfseries E-SPL\\\bfseries @1 m}
    & \makecell{\bfseries E-SPL\\\bfseries @3 m}
    & \makecell{\bfseries E-SPL\\\bfseries @5 m} \\
    \midrule

    \rowcolor{benchmarkgroupbg}
    \multicolumn{6}{l}{
      \strut\textbf{\emph{Open-source general-purpose}} models
    } \\

    InternVL2.5                     & 4B  & 3.24 & 1.33 & 0.68 & 0.70 \\
    InternVL2.5                     & 8B  & 2.92 & 1.29 & 0.67 & 0.60 \\
    InternVL2.5 INT8                & 38B & 2.98 & 1.23 & 0.62 & 0.63 \\
    InternVL2.5                     & 78B & 4.05 & 1.74 & 1.01 & 0.67 \\

    \addlinespace[2pt]

    Qwen2.5-VL                      & 7B  & 0.24 & 0.13 & 0.08 & 0.09 \\
    Qwen2.5-VL INT8                 & 32B & 0.15 & 0.05 & 0.02 & 0.03 \\
    Qwen2.5-VL BF16                 & 72B & 1.16 & 0.58 & 0.29 & 0.31 \\
    Qwen2.5-Omni visual-only        & 7B  & 1.73 & 0.71 & 0.37 & 0.34 \\
    Qwen3-VL-Instruct               & 8B  & 5.74 & 2.68 & 1.49 & 1.28 \\
    Qwen3-VL-Thinking         & 8B  & 2.22 & 1.11 & 0.65 & 0.49 \\

    \addlinespace[2pt]

    VideoLLaMA2                     & 7B  & 2.77 & 1.24 & 0.65 & 0.71 \\
    VideoLLaMA2                     & 72B & 4.68 & 2.00 & 0.99 & 0.95 \\
    VideoLLaMA2.1-16F               & 7B  & 2.59 & 1.13 & 0.53 & 0.62 \\
    VideoLLaMA2.1-AV visual-only    & 7B  & 7.76 & 3.43 & 1.90 & 1.54 \\

    \addlinespace[2pt]

    InternVideo2-Chat               & 8B & 1.17 & 0.51 & 0.25 & 0.31 \\
    LongVA-DPO                      & 7B & 3.50 & 1.65 & 0.99 & 0.89 \\
    MiniCPM-V2.6                    & 8B & 0.79 & 0.42 & 0.29 & 0.24 \\

     \rowcolor{benchmarkgroupbg}
    \multicolumn{6}{l}{
      \strut\textbf{\emph{Open-source emotion-recognition}} models
    } \\

    AffectGPT-frame                 & 7B & 3.31 & 1.32 & 0.63 & 0.74 \\
    Emotion-LLaMA                   & 7B & 7.88 & 3.03 & 1.75 & 1.43 \\

    Open-source mean (19)
    & -- & 3.10 & 1.35 & 0.73 & 0.66 \\

    \midrule

    \rowcolor{benchmarkgroupbg}
    \multicolumn{6}{l}{
      \strut\textbf{\emph{Closed-source general-purpose}} models
    } \\

    Gemini-3.0-Flash                & -- & 6.08 & 2.37 & 1.39 & 0.97 \\
    Gemini-3.0-Pro                  & -- & 9.11 & 3.93 & 2.34 & 1.52 \\
    GPT-5.2                         & -- & 4.23 & 1.75 & 1.05 & 0.78 \\
    GPT-5.6-sol                     & -- & 5.51 & 2.27 & 1.41 & 0.86 \\
    Claude Opus 5                   & -- & 11.27 & 4.83 & 2.66 & 2.03 \\

    Closed-source mean (5)
    & -- & 7.24 & 3.03 & 1.77 & 1.23 \\

    \bottomrule
  \end{tabular}
  \end{adjustbox}
\end{table}

\begin{figure}[t]
  \centering
  \includegraphics[
    width=0.88\linewidth
  ]{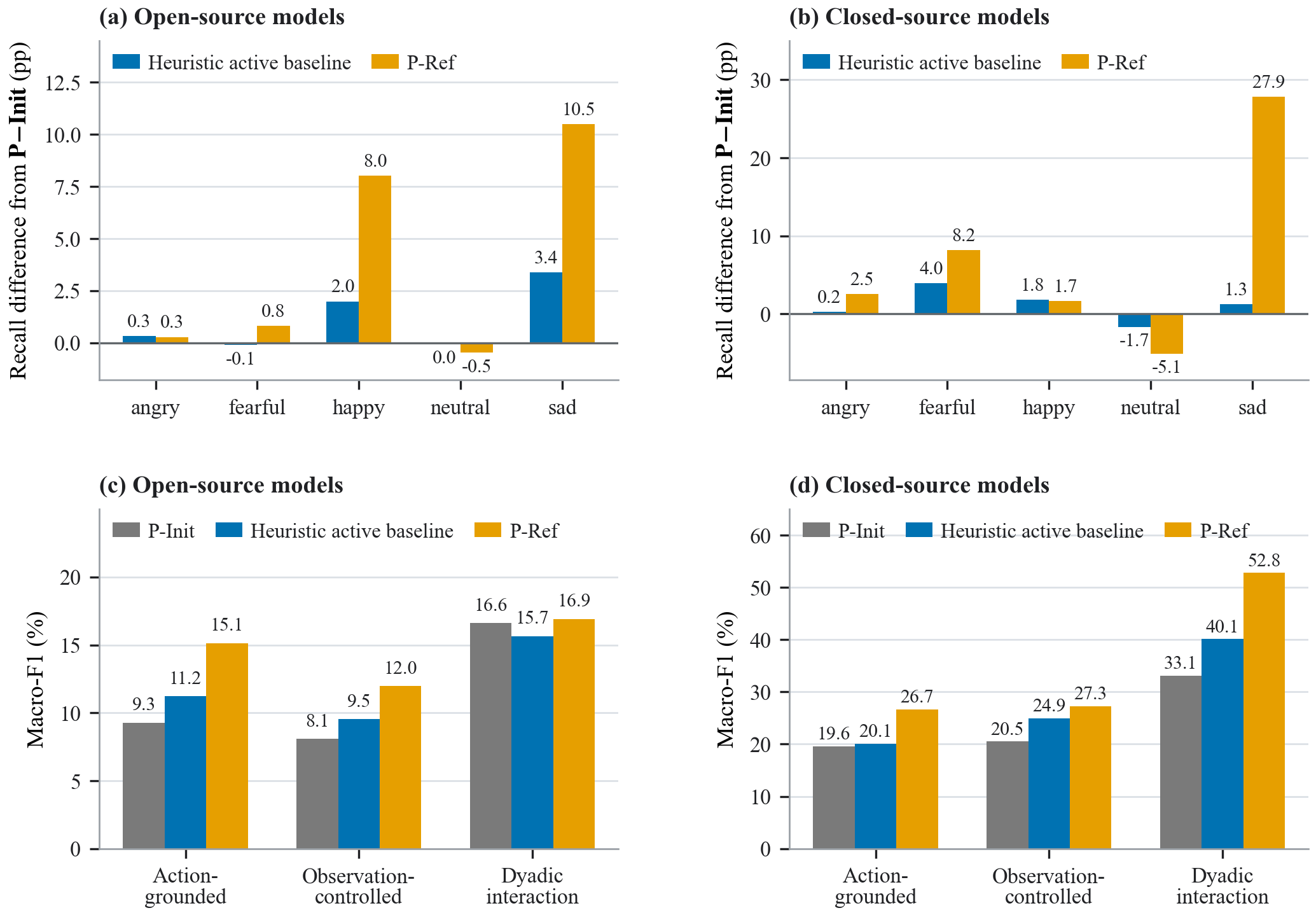}
  \caption{
     Class- and episode-family-level diagnostics on the complete
    3{,}941-episode test set.
    (a,b) Mean per-class recall gains of \textcolor{black}{heuristic baseline} and P-Ref relative to P-Init
    across the 19 open-source and five closed-source model configurations.
    (c,d) Mean Macro-F1 for action-grounded single-person
    ($n=2{,}899$), observation-controlled single-person ($n=832$), and
    dyadic-interaction ($n=210$) episodes. Each value is an unweighted mean
    of model-wise scores. 
  }
  \label{app:fig:class-action-diagnostics}
\end{figure}

\subsection{{Stopping-Time Controls and EAOG Comparison}}
\label{app:stopping-controls}
\textcolor{black}{The first legal fixed query time is 12\,s; the fixed-time controls displayed in main-paper Table~\ref{tab:eaot-timing-controls} query at 24, 30, or 48\,s. Uniform random samples one of the 37 integer candidate times from 12 through 48\,s. All use the same frozen learned observation policy and AffectSim-adapted, frozen VideoLLaMA2.1-7B-AV recognizer as EAOG, with 32-frame input on the same learned-system test split. The 24-s rule bisects the full 48-s budget; 30\,s bisects the legal candidate interval. Each method takes one recognition query from the preceding 12\,s of egocentric video and stops observation at its chosen time.}

\textcolor{black}{Main-paper Table~\ref{tab:eaot-timing-controls} reports Macro-F1, OGR, and E-SPL@3 for the displayed timing controls and EAOG. E-SPL@3 follows Eq.~\ref{eq:espl}; its eligible set is defined for the learned-system evaluation cohort separately from the zero-shot eligible counts in Table~\ref{app:tab:recovery-efficiency-full}. EAOG has the highest Macro-F1 and E-SPL@3 among the reported operating points.}

\textcolor{black}{Table~\ref{app:tab:eaot-method-ablation-full} reports P-Init and heuristic-baseline references, as well as accuracy, Half-Brier score, and mean termination time for the EAOG variants in main-paper Table~\ref{tab:eaot-loss-ablation}. The heuristic row uses rule-based viewpoint control after handoff and has no reported termination-time value. In this adapted-recognizer comparison, a failed handoff submits the available egocentric window; the zero-shot heuristic evaluation in Table~\ref{app:tab:main-results-full} uses paired P-Init fallback.}

\subsection{{Class and Episode-Family Diagnostics}}
\label{app:exp-variation}
\textcolor{black}{The lower panels compare P-Init, heuristic baseline, and P-Ref across action-grounded, observation-controlled, and dyadic-interaction episodes, using unweighted means of model-wise Macro-F1. Heuristic baseline improves the open-source action-grounded and observation-controlled means from 9.3\% to 11.2\% and from 8.1\% to 9.5\%, respectively, but lowers the dyadic mean from 16.6\% to 15.7\%. Closed-source heuristic baseline improves all three families, most strongly for dyadic interaction (33.1\% to 40.1\%); P-Ref reaches 52.8\% there. Thus, the aggregate gains do not imply uniform benefit across episode families.}

\begin{figure}[t]
  \centering
  \includegraphics[
    width=0.88\linewidth
  ]{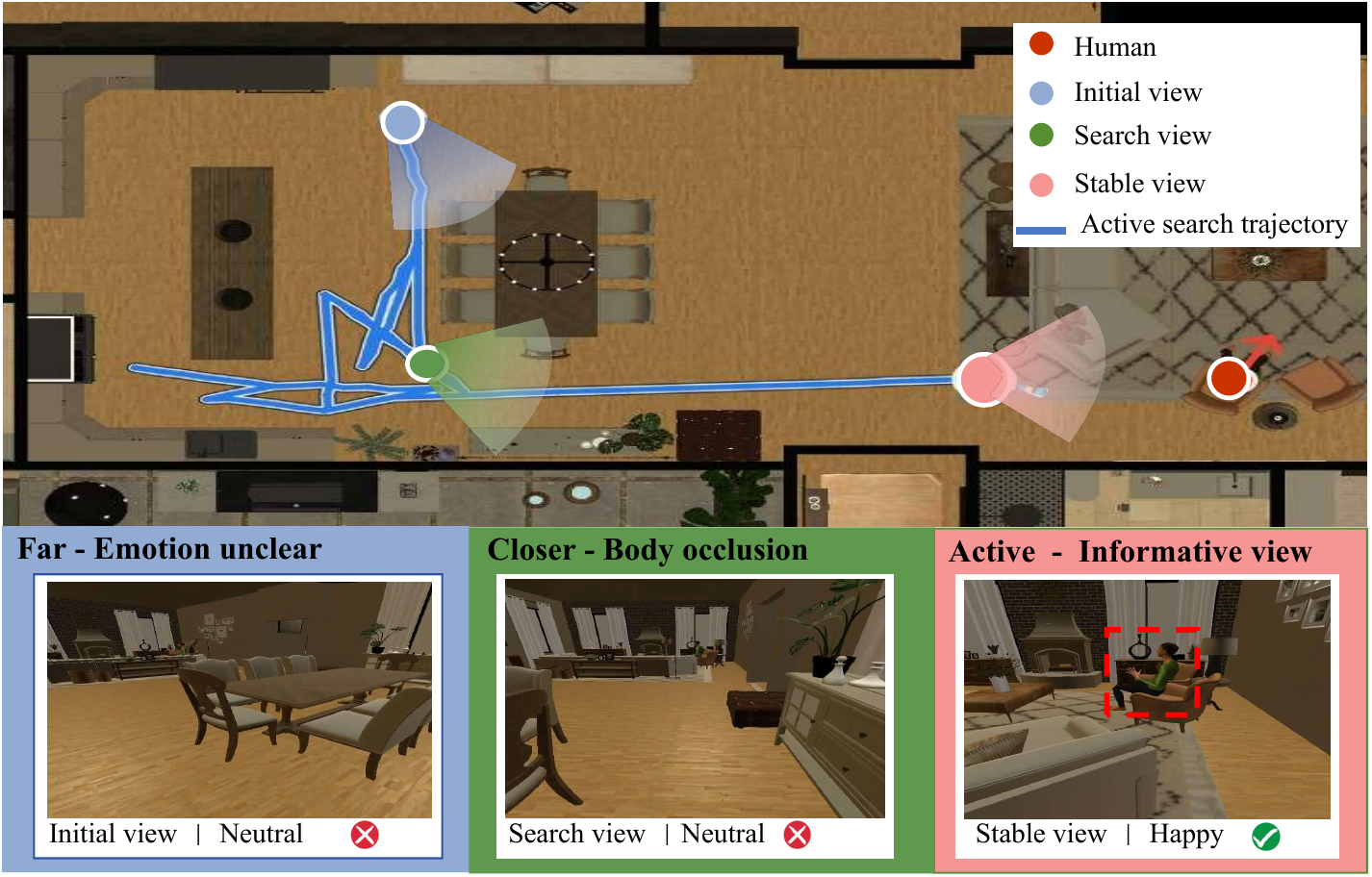}
\caption{
Example of active observation for emotion recognition. The top-down map shows
the performer, sampled robot viewpoints, and executed trajectory. The distant
initial observation is incorrectly recognized as neutral; moving closer still
leaves the body partially occluded, while further repositioning provides a
less-occluded observation that is correctly recognized as happy.
}
  \label{app:fig:active-observation-trajectories}
\end{figure}

\begin{table}[t]
  \centering
  \setlength{\belowcaptionskip}{\baselineskip}
  \caption{
    Execution diagnostics of \textcolor{black}{the heuristic baseline} on the complete
    test split. Mean path is computed per episode; handoff and budget
    exhaustion are percentages of episodes; pre-handoff path share is pooled
    over total locomotion; and clipped moves are normalized by attempted
    translations.
  }
  \label{app:tab:embodied-execution}

  \footnotesize
  \setlength{\tabcolsep}{5.5pt}
  \renewcommand{\arraystretch}{1.06}

  \begin{adjustbox}{max width=\textwidth}
  \begin{tabular}{cccccc}
    \toprule
    \textbf{Episodes}
    & \makecell{\textbf{Stable handoff}\\\textbf{(\% episodes)}}
    & \makecell{\textbf{Mean path}\\\textbf{(m/episode)}}
    & \makecell{\textbf{Path before handoff}\\\textbf{(\% locomotion)}}
    & \makecell{\textbf{Budget exhaustion}\\\textbf{(\% episodes)}}
    & \makecell{\textbf{Clipped moves}\\\textbf{(\% attempts)}} \\
    \midrule
    3{,}941 & 47.73 & 23.23 & 97.4 & 52.27 & 14.10 \\
    \bottomrule
  \end{tabular}
  \end{adjustbox}
\end{table}

\subsection{{Recovery, Efficiency, and Qualitative Example}}
\label{app:exp-recovery}
\textcolor{black}{Table~\ref{app:tab:recovery-efficiency-full} gives Recovery Rate and E-SPL for all 24 zero-shot recognizer configurations and their open- and closed-source group means. E-SPL evaluates recognition together with path efficiency relative to target regions at 1, 3, and 5\,m. These values use heuristic baseline and its eligible-episode counts, separately from the learned navigator evaluated in main-paper Table~\ref{tab:eaot-timing-controls}.}

\paragraph{\textbf{Episode-level recovery}}
Claude Opus 5 obtains the highest Recovery Rate at 11.27\%, while
Emotion-LLaMA achieves the highest value among the open-source
models at 7.88\%. These results provide an episode-level view complementary to the
aggregate Macro-F1 gains.

\paragraph{\textbf{Path-aware recognition}}
E-SPL evaluates correct \textcolor{black}{heuristic baseline} predictions together with the path efficiency of
observation acquisition. Across the 24 model configurations, mean E-SPL is
1.70\%, 0.95\%, and 0.78\% at target distances of 1, 3, and 5~m,
respectively. Claude Opus 5 obtains the highest E-SPL at all three distances,
while VideoLLaMA2.1-AV achieves the highest values among the open-source
models.

\textcolor{black}{Table~\ref{app:tab:embodied-execution} reports execution diagnostics for heuristic baseline, whose acquisition budget and fallback protocol differ from the learned-system evaluation.}

\textcolor{black}{Across all 3{,}941 test episodes, 47.73\% reach stable handoff and mean path length is 23.23\,m. Pre-handoff movement accounts for 97.4\% of locomotion, 52.27\% of episodes exhaust the original budget, and 14.10\% of translation attempts are clipped by the NavMesh.}

\textcolor{black}{Figure~\ref{app:fig:active-observation-trajectories} shows an episode in which repositioning reveals affective body-motion cues that approaching alone does not expose.}

\section{Limitations, Ethics, and Future Work}
\label{app:limitations}

\subsection{Limitations}

\benchmarkname{} has several limitations. First, the current benchmark relies
primarily on acted rather than spontaneous emotional expressions. Although
acted performances provide reproducible affective behaviors, they may not fully
capture the diversity and subtlety of naturally occurring emotion. Second,
avatar rendering and motion retargeting do not establish real-world transfer;
the current evaluation supports controlled within-benchmark conclusions about
observation and policy behavior. Third, the label space contains five discrete
emotion categories and does not represent mixed, continuous, or culturally
dependent affective states. Fourth, simulated humans follow predefined motions
and do not react to the observing agent. \benchmarkname{} therefore focuses on embodied
affective perception under controllable observation rather than fully
closed-loop affective interaction or deployment performance.

\subsection{Ethics and Broader Impact}

Emotion perception technologies raise important ethical concerns. The motion
capture protocol follows procedures for informed consent, compensation, data
governance, withdrawal, and data use. Released avatars do not reproduce the
participants' real facial appearance, reducing direct identity exposure
compared with releasing in-the-wild human video.

Nevertheless, emotion inference may be misused for surveillance, manipulation,
or psychological profiling. The benchmark is intended for research on embodied
perception, assistive robotics, and human-centered AI rather than high-stakes
decision making or surveillance. Release documentation will describe the
subjectivity and cultural limitations of emotion labels, intended uses, and
procedures for reporting potential misuse.

\subsection{Future Work}

Future versions may incorporate spontaneous affective behavior, finer-grained
or continuous affective states, multimodal cues, person-specific expression
styles, more diverse avatars and environments, and richer social
configurations. Reactive virtual humans would further allow the person's
behavior to change in response to the observer. Real-robot studies can then
test whether policies learned from the controlled benchmark retain their
observation strategies under real sensing and dynamics, treating transfer as a
separate empirical question rather than an assumption of the benchmark.

\end{document}